**Manuscript Title:** Resting-State Functional Connectivity Correlates of Emotional Memory Control under Cognitive load in Subclinical Anxiety

**Authors:** Shruti Kinger[1] and Mrinmoy Chakrabarty[1]

**Affiliations:** [1]Dept. of Social Sciences and Humanities, Indraprastha Institute of Information Technology Delhi, 110020, New Delhi, India

**Correspondence should be addressed to:** mrinmoy@iiitd.ac.in

**Number of Figures -** 7

**Number of Tables** - 3

Resting-State Functional Connectivity Correlates of Emotional Memory Control under Cognitive load in Subclinical Anxiety 1

**Abstract**

Volitional memory control supports adaptive cognition by enabling intentional Suppression of goal-irrelevant, interfering memories and Recall of goal-relevant memories. Neural mechanisms of Suppression and Recall have been studied largely in isolation, and their operation under concurrent working memory load in the context of subclinical anxiety remains unclear. We examined the control of emotionally valenced memories in 47 healthy participants with varying levels of subclinical anxiety under dual-task conditions, involving directed Suppression and Recall while concurrently performing a secondary task that imposed a visual working memory load. Cognitive efficiency in controlling dual-task memory-linked interference, measured by the Balanced Integration Score (BIS), showed no differences between Suppression vs. Recall; across emotions or influence from anxiety. Intrinsic functional brain networks measured by seed-to-voxel resting-state functional connectivity (rsFC) revealed dissociable rsFC profiles linked to Cognitive Control across emotional valences, and these relationships were moderated by anxiety. Efficient Suppression of positive memories correlated with reduced connectivity between anterior cingulate cortex and posterior perceptual-midline regions, and diminished hippocampal-frontal pole coupling. Conversely, efficient Suppression of negative memories correlated with increased posterior parietal - lateral occipital regional connectivity. Anxiety particularly moderated associations between Cognitive Control and prefrontal connectivity during Suppression of positive memories and Recall of positive and neutral memories. Direct comparisons further revealed stronger hippocampal–thalamic rsFC during Suppression vs. Recall of positive memories. Together, we delineate the neural correlates of volitional emotional memory control under cognitive load and suggest that subclinical anxiety severity shapes these networks selectively, providing targets for future mechanistic studies.

## 1. Introduction

Flexibly regulating memory representations through intentional recall or suppression is fundamental to adaptive cognition. Volitional recall supports goal-directed retrieval, whereas volitional suppression prevents undesirable memories from intruding into conscious awareness, thereby insulating ongoing cognitive processes (Anderson & Hanslmayr, 2014). Together, these complementary mechanisms enable optimal allocation of limited cognitive resources in meeting the demands of dynamic environments.

Evidence suggests that recall and suppression rely on partially overlapping but functionally distinct mechanisms. Directed recall is associated with enhanced hippocampal engagement and coordinated activity within default mode and frontoparietal networks that support episodic retrieval and maintenance of internal representations (Moscovitch et al., 2016; Rizio & Dennis, 2013). Conversely, voluntary suppression is an effortful, active process mediated by top-down control from prefrontal regions, particularly the dorsolateral and inferior frontal cortices which downregulate hippocampal activity to constrain retrieval (Anderson et al., 2016; Bastin et al., 2012; Wylie et al., 2008). These convey the active nature of control operations which compete for executive resources.

Failures of memory control can impair cognitive performance. Ineffective suppression allows intrusive memories to disrupt attentional control and working memory (Eysenck et al., 2007; Muñoz et al., 2013). Similarly, excessive engagement in recall, particularly of emotionally salient material may overburden cognitive resources and interfere with concurrent task demands. Resting-state functional connectivity (rsFC) studies further implicate large-scale networks linking frontal control regions with temporal and midline structures involved in internally generated thought in the regulation of intrusive cognition (Lu et al., 2022). Such disruptions in regulatory control are especially pronounced in anxiety.

Anxiety may present either as a clinical anxiety disorder, requiring specific symptom counts, duration, and impairment as defined by DSM-5-TR/ICD-11 criteria (American Psychiatric Association [APA], 2013; World Health Organization [WHO], 2024) or as subclinical (subthreshold) anxiety, characterized by elevated symptoms that do not meet the full criteria for clinical diagnosis (Volz et al., 2022). This distinction is particularly relevant in light of proposals advocating a dimensional approach, in which anxiety is conceptualized along a continuum of symptom severity (Brown & Barlow, 2009; Okasha, 2009; Volz et al., 2022). Although individuals with subclinical anxiety typically experience milder symptoms and less functional impairment than those with clinically diagnosed anxiety, subthreshold presentations are more prevalent and remain clinically meaningful (Volz et al., 2022; Witlox et al., 2021; Zhong et al., 2024). In the present study, we focus specifically on subclinical anxiety, defined as variation in self-reported anxiety symptom severity within a general young adult population that does not meet diagnostic criteria for an anxiety disorder but nonetheless influences cognition, emotional processing and behaviour. This represents an important at-risk stage that may precede the development of clinically significant anxiety (Bishop, 2009; Eysenck et al., 2007), making it relevant to characterize underlying behavioural and neural patterns early. Accordingly, throughout this manuscript, the term "anxiety" refers to subclinical anxiety unless otherwise specified, consistent with a dimensional framework that distinguishes it from clinically diagnosed anxiety disorders.

Subclinical anxiety has been consistently associated with altered memory control and inefficient allocation of cognitive resources aside from influencing other higher cognitive functions as follows. Individuals with elevated anxiety show increased susceptibility to intrusive emotional memories, reduced suppression efficacy, and enhanced recall of negative memory (Dieler et al., 2014; Marzi et al., 2014; Waldhauser et al., 2011). At the network level, anxiety is linked to heightened engagement of frontoparietal, posterior cingulate, and precuneus regions, reflecting increased self-referential processing and perseverative cognition (Bijsterbosch et al., 2014). Behavioural evidence further suggests that anxiety compromises sensory information processing (Achyuthanand et al., 2023; Chakrabarty et al., 2021; Kaur et al., 2023) and working memory efficiency under high cognitive demands by depleting central executive capacity (Bishop, 2009; Eysenck et al., 2007).

In this backdrop, relatively little is known about the interaction of these processes with competing cognitive load, particularly in individuals with subclinical anxiety. In everyday contexts, memory control seldom occurs in isolation; rather, individuals must regulate memory while simultaneously performing other cognitively demanding tasks. Cognitive load reduces available working memory resources, making both Suppression (Wang et al., 2020; Wegner et al., 1993) and Recall more difficult and increasing susceptibility to compromises. If Recall and Suppression impose differential demands on executive control, these differences should be reflected in performance on an independent working memory task and may be further modulated by anxiety. Moreover, while task-based neuroimaging has clarified the neural substrates of memory control, associations between individual differences in emotional memory control, cognitive efficiency, and rsFC remain insufficiently characterized.

To address these gaps, we employed an item-method directed forgetting paradigm to examine how directed Recall and Suppression of emotionally valenced stimuli are influenced with competing visual working memory task, indexing Cognitive Control through (dual-task) memory-linked interference. We further examined how individual differences in Cognitive Control and subclinical anxiety relate to rsFC within large-scale neural networks implicated in memory, emotion, and executive control. Specifically, we examined a) whether dual-task memory-linked interference / Cognitive Control varies across emotional valences during Recall and Suppression; b) whether rsFC is associated with dual-task memory-linked interference / Cognitive Control across valences in both conditions; and c) whether anxiety influences behaviour and moderates brain-behaviour associations. Along the lines of these goals, we hypothesised a) difference in dual-task memory-linked interference / Cognitive Control between Recall and Suppression across emotional valences; b) association between dual-task memory-linked interference / Cognitive Control and rsFC across valences in both conditions; and c) anxiety influences dual-task memory-linked interference / Cognitive Control for both positive and negative emotional valences across both conditions, and it moderates the association between brain and behaviour. Here, we provide evidence supporting the latter two questions related to network-level functional patterns of volitional memory control under competing cognitive demands and discuss the absence of effects for the first.

## 2. Methods

### 2.1. Participants

Participants were recruited through emails and noticeboard postings on the university campus, as well as via email outreach to individuals outside the campus. Forty-seven participants (14 females) were included based on the following inclusion criteria, (1) 18-35 years of age, (2) normal or corrected vision and no colour blindness,

(3) no reported diagnosis of neurological/psychiatric disorder in the last three months with no history of epilepsy, (4) not on any prescription drugs for the nervous system and/or respiration (asthma), and (5) consent to participate in both the behavioural and magnetic resonance imaging (MRI) experiment. [see Table 1 for demographic details]

Our sample size estimation was informed by four prior studies closely aligned with our research questions and experimental design. The effect sizes from these studies were harmonized by converting the reported statistics to Pearson's *r* following steps mentioned elsewhere (Cohen, 1988). First, Fawcett and Taylor (2008) reported that memory suppression imposes greater cognitive demands than intentional recall in a sample of n = 25 participants, yielding a large effect (*Cohen's f* = 0.82), which corresponds to *r* = 0.63. Second, a behavioural study (Dieler et al., 2014) investigating thought suppression (n = 36) demonstrated that higher anxiety was associated with poorer suppression performance for negatively valenced stimuli (*r* = 0.45). Third, in a sample of n = 47 participants (Waldhauser et al., 2011), elevated trait anxiety predicted reduced suppression efficacy, indexed by greater memory strengthening across repeated suppression attempts (β = 0.49, equivalent to *r* = 0.49). Finally, a neuroimaging study (Nowicka et al., 2011) in a sample of n = 16 participants examining emotional influences on directed forgetting reported significantly greater neural activation during forgetting compared with recall for negative images (lowest *t* = 4.09 corresponding to *r* = 0.71). To derive a conservative estimate, we used the smallest effect size reported across the above studies, i.e., *r* = 0.45, in the tool G*Power (version 3.1.9.7; Faul et al., 2007), specifying an Exact test for correlation (bivariate normal model), a two-tailed significance level of α = 0.05, and desired power of 0.80. This yielded a minimum required sample size of 36 participants. We here report our results with a final sample of 47 participants which is beyond the minimum required sample to ensure adequate statistical power. The study and its procedures were performed in compliance with laws and institutional guidelines and were approved by the Institutional Ethics / Review Board of Indraprastha Institute of Information Technology Delhi, INDIA vide letter number EC/NEW/INST/2024/DL/0440.

### 2.2. Subjective Ratings

Self-report measurements were collected using PHQ-9 (Patient Health Questionnaire; Kroenke et al., 2001) to screen for depression and participants with PHQ ≤ 15 were included in the analysis, and Fear Affect scale of National Institutes of Health toolbox to measure the cognitive component of anxiety to assess self-reported fear and anxious misery (Pilkonis et al., 2013). The participants were provided with instructions to rate the items as accurately as they can. They rated their feelings of fear and anxiety over the past seven days on the Fear Affect scale using a 5-point Likert scale (1 = never to 5 = always). Although the possible score range for this scale is 7 to 35, the observed scores in the present sample fell within a 14 ± 7 range. Total scores were computed by summing item responses. In the absence of culture-specific normative data, the median score of the sample was used to classify participants into two subsets: a "relatively low-anxiety subset" (scores ≤ median) and a "relatively high-anxiety subset" (scores > median). The behavioural experiment commenced after completion of the questionnaires.

### 2.3. Behavioural experiment:

We employed two behavioural experiments- item method directed forgetting task under a concurrent visual working-memory (WM) load (Figure 1) and a recognition task on day 1, followed by a separate resting state functional magnetic resonance imaging (rs-fMRI) on day 2. The time interval between day 1 and 2 was 16 ± 15 days**.** Both the behavioural experiments were designed using Psychopy and participants were seated 57 cm away from a 24-inch monitor. A chinrest was used to stabilise participant's head. The experiment began with a fixation point for 1000 ms followed by an image (subtending visual angle of ~14° × 10° or ~10° × 14°) of negative, neutral, and positive valence for another 1000 ms. The instructions were to view the image. A blank screen appeared after the images for a duration of 1000 ms which was followed by a red or green fixation point indicating task instruction. If the fixation point appeared in red, the instruction imparted was to Suppress the image and in case of a green colour fixation point, the instruction was to Recall the image in their memory. The blank interval of 1000 ms following the image was to account for volitional control to kick in (Bengson et al., 2015). After this, the independent visual working memory task followed where the participant was asked to press the right key if the orientation of the second stimulus was tilted clockwise with respect to the first one and press left key if the tilt was to the left or anticlockwise. The orientation of the first Gabor patch was chosen between 60° and 300°. The orientation of the second Gabor patch was tilted relative to the first by one of the following angles: ± 4°, ± 5°, ± 7°, ± 9°, ± 12°, ± 15°, ± 20°, ± 26°, ± 34°, or ± 45°. The stimulus (Gaussian Gabor patch) had a size of 3 degrees, a spatial frequency of 3 cycles/degree, and a contrast of 100%. Each participant completed three sessions, with each session containing 60 trials. Each session included 30 Recall trials and 30 Suppress trials, further divided into 10 trials each for negative, neutral, and positive valence images. The images were sourced from NAPS database (Marchewka et al., 2014) and were validated (n = 15) separately on a 9-point rating scale for both valence and arousal. The score of 1 on a valence scale indicated extremely unpleasant and a score of 9 indicated extremely pleasant. The score of 1 on the arousal scale indicated extremely bored and score of 9 indicated extremely excited. There was significant difference for both valence (negative: 2.30 ± 0.38; neutral: 5.09 ± 0.20; positive: 7.67 ± 2.67) and arousal (negative: 6.77 ± 0.59; neutral: 5.13 ± 0.49; positive: 3.84 ± 0.81) in line with NAPS database.

Thereafter, a recognition task was administered to assess participants' adherence to the Recall and Suppress instructions where participants were asked to respond whether the images presented were old or new. The task comprised 270 images in total with 180 images that were presented earlier in the three sessions and 90 foils (negative = 30; neutral = 30, positive = 30).

------------------------------------------------------------------------------------------------------------------------------------

Insert Figure 1

------------------------------------------------------------------------------------------------------------------------------------

**2.4. Behavioural data analyses**

To this end, median reaction times (RT; computed from correct trials) and accuracy (defined as the proportion of at least 33 correct responses out of 60 trials per session) were derived for each of the six conditions: negative Recall, neutral Recall, positive Recall, negative Suppress, neutral Suppress, and positive Suppress. From these measures, we calculated a composite index of dual-task memory-linked interference / Cognitive Control - Balance Integration Score (BIS) (Madrid and Hout, 2019). BIS was computed by first calculating the z-scores of RT and accuracy across the six conditions for each participant followed by subtracting the z-scored median RT from accuracy as explained in Equation 1, such that higher values indicate lesser memory-linked interference or better Cognitive Control. The resulting BIS scores are presented in Figure 2A.

$BIS = zscore_{accuracy} - zscore_{RT}$ (Equation 1)

The metric represents interference under competing demands for limited executive control related cognitive resources. Any un-regulated / uncontrolled memory representation would leak into the concurrent working memory (WM) task-engaged demands and interfere with it. It is noteworthy that interference stems from successful retrieval of memories during the Recall condition and from failed suppression of memories during the Suppress condition. Thus, with regards to our study the balance integration score or BIS indexes memory-related processes interfere with concurrent task performance. Higher BIS on the visual working memory task was interpreted as reflecting reduced activation or interference of memory representations. In the Suppress condition, this reduction applied to goal-irrelevant memories, indicating downregulation via cognitive control, which in turn minimized interference and supported improved WM performance. In contrast, in the Recall condition, higher BIS reflected reduced activation or interference of goal-relevant memories, likely due to ineffective retrieval. Previous research suggests that cognitive control exerted through suppression reduces interference by downregulating memory representations (Festini & Reuter-Lorenz, 2014). Moreover, dual-task costs have been shown to impair performance (Hensen et al., 2024) therefore maintaining an active memory during Recall may be detrimental to a concurrent visual working memory task. Thus, a higher BIS reflects greater cognitive control through reduced memory interference during concurrent working memory tasks. The validity of these interpretations was evaluated through separate confirmatory analyses (see Supplementary section S1).

To ensure adherence to task instructions, we measured performance on a recognition task and computed sensitivity scores for directed Suppression and Recall instructions. Higher sensitivity scores indicate greater recognition of images that were either recalled or suppressed in accordance with task instructions. We found that the sensitivity scores were significantly larger for the Recall task instruction compared to the Suppression task instruction. This difference in memory performance between the two conditions indicated successful instruction compliance.

Sensitivity scores were computed for the recognition task to quantify differences in directed remembering versus forgetting for each emotional category and averaged across emotions separately for each instruction type. The aggregate sensitivity scores for each task instruction are presented in Figure 2B. All data and statistical analyses were done using custom code written in MATLAB (R2022b).

------------------------------------------------------------------------------------------------------------------------------------

Insert Figure 2

------------------------------------------------------------------------------------------------------------------------------------

**2.5. MRI experiment**

Scans were acquired on a 32-channel head coil 3 Tesla Signa Architect (GE). The imaging parameters for the T1-weighted anatomical images were as follows, voxel size= 0.5 × 0.5 × 0.5 $mm^3$, slice count = 344 slices, TR (repetition time) = 2,500 ms, TE (echo time) = 2.4 ms, flip Angle = 25°, matrix: 512 × 512. T2- weighted functional scans were acquired for a duration of ~11 minutes (eyes open) using echo-planar imaging (EPI) sequence with the following parameters: TR = 2,000 ms, TE = 30 ms, flip angle = 90°, resolution matrix= 64 × 64, field of view (FOV)= 200 × 200 $mm^2$, slice count = 33 slices, and voxel size: 3.1 × 3.1 × 5 $mm^3$. Resting-state fMRI data were collected in a single session at Mahajan Imaging Labs, SDA, New Delhi, and the acquisition parameters were consistent with those reported in previous study (Zhang et al., 2020).

The preprocessing in CONN toolbox (version 22a; (Whitfield-Gabrieli & Nieto-Castanon, 2012)) implemented in MATLAB (version R2024a) was carried out using the default pre-processing pipeline which comprised of the following steps- 1) functional realignment and unwarping was done after correcting for head motion, 2) slice-timing correction, 3) outlier detection (framewise displacement > 0.9mm; global signal ± 5 SD),

4) direct segmentation and normalization of both functional & structural data in standard MNI space, and 5) smoothing performed on functional scans with a Gaussian kernel window of 8 mm full width half maximum. After pre-processing, the confounding effects of initial few scans, cerebrospinal fluid and white matter, and motion were removed in denoising and temporal band-pass filtering was performed to remove frequencies below 0.008 or above 0.09 Hz as part of denoising step. Mean head motion observed across participants was low (mean = 0.10 mm, standard deviation = 0.04 mm), indicating minimal subject movement during scanning. Subsequently, seed-to-voxel functional connectivity was computed by correlating the blood-oxygenation-level-dependent (BOLD) time series of the seed region of interest (ROI) with the BOLD time series of every voxel in the brain for each participant.

A priori regions of interest (ROIs) to be set as seeds for statistical analyses were defined using a structured, multi-step procedure. First, we conducted targeted searches on PubMed using hypothesis-relevant keywords to identify peer-reviewed journal studies (published between 2000 and 2025) addressing the constructs central to each hypothesis. Second, we screened the resulting literature for empirical neuroimaging studies that were directly relevant to each hypothesis and methodologically comparable to our design but found no suitable results. Therefore, we depended on systematic review to choose a priori ROIs separately for each hypothesis. Third, to validate and refine these candidate regions, we further conducted automated meta-analytic checks using the NeuroQuery tool (Dockès et al., 2020) for overlapping regions based on the weight of the cognitive construct on the brain map generated using the tool. Finally, for each hypothesis, the final set of ROIs comprised the three regions most consistently implicated across the literature, selected based on convergent empirical evidence and neurobiological plausibility, as detailed below. The hypotheses, ROIs, search keywords and NeuroQuery weights are in Table 2. This approach ensured that ROI selection was firmly grounded in prior research. Restricting each hypothesis to three a priori ROIs served both theoretical and statistical aims - it limited the number of planned comparisons; reduced Type I error inflation after multiple-comparison correction and maintained statistical power given the modest sample size (n = 47). Constraining the ROI set in advance also ensured that statistical tests remained focused, interpretable, and anchored in established neurobiological frameworks.

Consequently, the following brain regions were selected as ROIs, from Harvard Oxford Atlas and HCP Glasser Atlas (Desikan et al., 2006) in MNI space, along with their corresponding neurobiological rationale: (a) Anterior Cingulate Cortex as it had been implicated in regulating control over the dorsolateral prefrontal cortex and hippocampus for managing intrusive thoughts (Anderson & Hanslmayr, 2014; Crespo-García et al., 2022; Rolls, 2019; Shackman et al., 2011), (b) Hippocampus for its critical role in memory encoding, retrieval and suppression (Anderson et al., 2025; Anderson & Hanslmayr, 2014; Crespo-García et al., 2022; Moscovitch et al., 2016; Simons & Spiers, 2003), (c) Supramarginal gyrus as it has been associated with episodic memory and in directing attention to task relevant goals, and emotion processing in anxiety (Cabeza et al., 2008; Fonzo et al, 2015; Seghier, 2013; ), (d) Precuneus, associated with mental imagery, episodic memory retrieval, and memory suppression in both healthy and trauma-exposed individuals (Mary et al., 2020; Rizio & Dennis, 2013), (e) Amygdala, involved in modulation of emotional memory (Benarroch, 2014) and anxiety (Bijsterbosch et al., 2014; Bishop, 2009) , (f) Middle frontal gyrus, implicated in inhibition, emotion regulation, and its involvement in exercising anxiety modulation (Anderson & Hanslmayr, 2014a; Navarro-Nolasco et al., 2025), (g) Superior frontal gyrus for its involvement in emotional information processing, inhibitory control, and anxiety (Molent et al., 2018; Tomasino et al., 2024), (h) pars triangularis Inferior frontal gyrus, as it plays a crucial role in inhibition and assigning meaning to stimulus especially in anxiety (Depue et al., 2007; Sullivan et al., 2019; Swick et al., 2008). Except anterior cingulate cortex, all ROIs were anatomically defined.

**2.6. MRI data analyses**

Using the CONN toolbox, first-level analysis was performed on pre-processed and denoised rs-fMRI data to compute subject-specific functional connectivity measures of seed-to-voxel correlations. These individual connectivity maps were then entered into second-level analyses, where group-level statistical tests were conducted to identify significant connectivity patterns and between-subject effects. All analyses were performed using bivariate correlation models in the CONN toolbox, with age, gender and delay (time interval) between day 1 (behavioural experiment) and day 2 (MRI experiment) included as nuisance covariates (see Equations 2–4). First, we assessed the association between resting-state functional connectivity (rsFC) and dual-task memory-linked interference / Cognitive Control (BIS) – Suppression and Recall (Equation 2). Second, we tested whether this association was moderated by anxiety severity by including an BIS × anxiety interaction term (Equation 3). Third, differences in rsFC associated with dual-task memory-linked interference / Cognitive Control for Recall versus Suppression were examined across the three emotional valences separately (Equation 4).

No outliers (± 3 SD from the mean) were detected. Statistical inference was based on a voxel-wise threshold of $p < 0.001$ (uncorrected) and a cluster-level threshold of $p < 0.05$ (FWE-corrected). Results were visualized using MRICroGL (Rorden & Brett, 2000), and anatomical labels assigned using the Harvard–Oxford Atlas and functional label using HCP Glasser Atlas (Desikan et al., 2006). Seed regions (shown in green) and thresholded statistical maps, adjusted for the number of *a priori* regions per hypothesis ($p$FWE $< 0.05/3 \approx 0.017$; see Table 3), were overlaid on a group-level structural template generated from the averaged, pre-processed, unsmoothed T1-weighted images of all 47 participants using SPM version 12 (https://www.fil.ion.ucl.ac.uk/spm/software/spm12/).

For better interpretation of the interaction between the dual-task memory-linked interference / Cognitive Control and anxiety in explaining rsFC, we visualized the effects using scatter plots with fitted trend lines, after splitting the participants by median ratings on anxiety scale (median ± interquartile range [iqr] = 14 ± 7) into ' relatively-high' (> 14; n = 23) and 'relatively-low' anxiety (≤ 14; n = 24) subsets (see MRI results).

$$y = \beta 0 + \beta 1(age) + \beta 2(gender) + \beta 3(delay) + \beta 4(BIS) + \epsilon \quad \text{(Equation 2)}$$

$$y = \beta 0 + \beta 1(age) + \beta 2(gender) + \beta 3(delay) + \beta 4(BIS) + \beta 5\,(Fear\ Affect) + \beta 6(BIS \times Fear\ Affect) + \epsilon \quad \text{(Equation 3)}$$

$$y = \beta 0 + \beta 1(age) + \beta 2(gender) + \beta 3(delay) + \beta 7(BIS{:}Suppression) + \beta 8(BIS{:}Recall) + \epsilon \quad \text{(Equation 4)}$$

where y= rsFC between seed / ROI and voxel; β0 = intercept; β1-β8= parameter estimates and ϵ= residual. Age, gender, and delay between behavioural experiment and MRI experiment (in number of days) were covariates of no interest in all the three models (Equations 2-4). The first regression model (Equation 2) measured the effect of dual-task memory-linked interference / Cognitive Control following directed Suppression or Recall instruction on rsFC; the second regression model (Equation 3) measured the interaction between dual-task memory-linked interference / Cognitive Control (Suppression / Recall) and Fear Affect (please note that from here on, we will use the term 'anxiety' for Fear Affect) on rsFC after controlling for the main effects of dual-task memory-linked interference / Cognitive Control and anxiety besides controlling for the nuisance covariates; the third model (Equation 4) compared dual-task memory-linked interference / Cognitive Control between 'Suppression' condition and 'Recall' condition. The independent variables in the model were orthogonal.

## 3. Results

### 3.1. Behavioural results

To examine the effects of task instruction and emotional valence, we conducted a two-way repeated-measures analysis of variance (ANOVA) with task instruction (Recall, Suppress) and emotion (negative, neutral, positive) as within-subject factors and dual-task memory-linked interference / Cognitive Control as dependent variable. Assumptions of normality (Lilliefors test; all *ps* > 0.40) and homogeneity of variance (Mauchly's test for sphericity; $W$ = 0.68, $p$ = 0.18) were met. Besides, no outliers above or below 2.5 standard deviations were identified. The analysis revealed no significant main effect of task instruction ($F_{(1,46)}$ = 0.05, $p$ = 0.82, partial $\eta^2$ = 0.001), emotion ($F_{(2,92)}$ = 2.55, $p$ = 0.08, partial $\eta^2$ = 0.05), and no interaction between task instruction and emotion ($F_{(2,92)}$ = 0.98, $p$ = 0.38, partial $\eta^2$ = 0.02). These results indicate that imposing a concurrent visual WM load while participants engaged in Recall or Suppression of emotional memories did not result in differences in dual-task memory-linked interference / Cognitive Control. We had hypothesised that if Suppression were more effortful than Recall (or vice versa), the more demanding process would recruit greater cognitive resources and thereby differentially impact dual-task memory-linked interference / Cognitive Control. The absence of behavioural differences may suggest that Recall and Suppression elicited comparable working memory engagement, i.e., participants exhibited similar working memory capacity across conditions. We further computed correlation between anxiety and dual-task memory-linked interference / Cognitive Control for each of the six conditions reported above and found all *p*s > 0.29.

To ensure compliance for task instructions in a recognition task (see Methods), a paired t-test (Lilliefor's test for normality; all *p*s > 0.1) revealed a significant difference between the sensitivity scores of Recall (mean ± s.e.m. = 1.55 ± 0.51) and Suppress (mean ± s.e.m. = 1.42 ± 0.51) conditions ($t_{(46)}$ = 2.39, $p$ = 0.02, *Cohen's d* =

0.34), indicating that participants correctly recognised more images in the Recall than in the Suppress condition.

### 3.2. MRI results

#### 3.2.1. *Associations Between Suppression-linked Cognitive Control of Emotional Memories and Resting-State Functional Connectivity*

Suppression-linked Cognitive Control of positive memories, showed a significant negative association with rsFC between the anterior cingulate cortex (ACC) seed of the salience network and a posterior cluster encompassing the intracalcarine cortex (ICC), precuneus, and supracalcarine cortex cortex (SCC; $T_{(42)}$ = -5.06, *p*FWE = 0.02, cluster size *kE* = 246, peak at x = +10, y = -66, z= +18; Figure 3A). In addition, a significant negative correlation was observed between the right hippocampal seed and a cluster in the frontal pole (FP; $T_{(42)}$ = -5.69, *p*FWE = 0.006, cluster size *kE* = 317, peak at x = +22, y = +56, z = +12; Figure 3C). To illustrate the direction and magnitude of the effects, rsFC values extracted from this cluster showed a negative association with Suppression-linked Cognitive Control of positive memories or BIS (Figure 3B: $r_{(45)}$ = -0.55, $\beta$ = -0.09, 95% *CI* [-0.73, -0.32]; Figure 3D: $r_{(45)}$ = -0.65, $\beta$ = -0.10, 95% *CI* [-0.79, -0.44]). These inverse relationships indicate that greater Suppression-linked Cognitive Control was associated with reduced functional coupling between these seed regions and their respective target clusters, consistent with engagement of inhibitory control networks during memory suppression (Harita et al., 2024; Qian et al., 2018).

A distinct network was identified for Suppression-linked Cognitive Control of negative memories. Specifically, the Suppression-linked Cognitive Control was positively associated with rsFC between the right posterior supramarginal gyrus (pSMG) seed and a cluster encompassing the left angular gyrus and superior division of the lateral occipital cortex ($T_{(42)}$ = 4.64, *p*FWE = 0.017, cluster size *kE* = 261, peak at *x* = −52, *y* = −58, *z* = +22; Figure 3E). To illustrate the direction and magnitude of the effect, rsFC values extracted from this cluster showed a positive association with Suppression-linked Cognitive Control of negative memories or BIS (Figure 3F: $r_{(45)}$ = 0.55, $\beta$ =0.16, 95% *CI* [0.32, 0.72]). This positive association suggests that stronger functional coupling within this network supports more efficient suppression of negative memories (Qian et al., 2018). Together, these findings indicate that Suppression-linked Cognitive Control for negative memories scaled with increased resting-state functional connectivity between the pSMG and posterior representational regions.

No resting-state functional connectivity network showed a significant association with Suppression-linked Cognitive Control of neutral memories.

---

Insert Figure 3

---

### 3.2.2. *Associations Between Recall-linked Cognitive Control of Emotional Memories and Resting-State Functional Connectivity*

No association between Recall-linked Cognitive Control and rsFC survived the stringent family-wise error (FWE) correction.

### 3.2.3. *Anxiety-Moderated Associations Between Emotional Memory Suppression-linked Cognitive Control and Resting-State Functional Connectivity*

The interaction between anxiety and Suppression-linked Cognitive Control of positive memories revealed a significant positive association with rsFC between the right superior frontal gyrus (SFG r) seed and a cluster encompassing the medial frontal cortex (MedFC), left paracingulate gyrus (PaCiG), and anterior cingulate gyrus (AC; $T_{(40)} = 5.81$, *p*FWE < 0.001, cluster size *kE* = 617 with peak at x = -04, y = +44, z = -12; Figure 4A). Two separate visualizations revealed a positive correlation in the high (Figure 4B: $r_{(21)} = 0.58$, $\beta = 0.09$, 95% *CI* [0.22, 0.80]) and negative correlation in the low (Figure 4B: $r_{(22)} = -0.51$, $\beta = -0.07$, 95% *CI* [-0.76, -0.14]) subset between the Suppression-linked Cognitive Control of positive memories or BIS and seed-to-voxel rsFC. Closer inspection of the data confirmed the difference between the two slopes, which suggests that the rate of change in rsFC with unit increment of BIS was higher in the high anxiety subset as compared to the low anxiety subset.

The moderating effect of anxiety on Suppression-linked Cognitive Control for negative memories showed trends under more lenient correction thresholds but did not survive the more stringent family-wise error (FWE) correction. Full details of these analyses are provided in the Supplementary Materials (S3).

---------------------------------------------------------------------------------------------------------------------------------

Insert Figure 4

---------------------------------------------------------------------------------------------------------------------------------

### 3.2.4. *Anxiety-Moderated Associations Between Emotional Memory Recall-linked Cognitive Control and Resting-State Functional Connectivity*

The interaction between anxiety and Recall-linked Cognitive Control of positive memories revealed a significant negative association with rsFC between the left inferior frontal gyrus pars triangularis (IFG tri l) seed and three target clusters. Cluster 1 included the right temporo-occipital middle temporal gyrus (toMTG r; $T_{(40)} = -5.59$, *p*FWE = 0.008, cluster size *kE* = 270, peak at *x* = +66, *y* = −50, *z* = −10; Figure 5A) and cluster 2 included the right inferior frontal gyrus pars opercularis (IFG oper r; $T_{(40)} = -5.59$, *p*FWE = 0.012, cluster size *kE* = 270, peak at *x* = +46, *y* = +20, *z* = +18; Figure 5B). Different visualizations conducted separately within anxiety subsets for the three target clusters revealed distinct association patterns between Recall-linked Cognitive Control of positive memories or BIS and rsFC. Cluster 1 showed a negative correlation in the high (Figure 5B: $r_{(21)} = -0.31$, $\beta = -0.05$, 95% *CI* [-0.64, 0.11]) and a positive correlation in the low (Figure 5B: $r_{(22)} = 0.49$, $\beta = 0.09$, 95% *CI* [0.10, 0.74]) anxiety subset. Cluster 2 exhibited a negative correlation in the high (Figure 5D: $r_{(21)} = -0.35$, $\beta = -0.05$, 95% *CI* [-0.66, 0.08]) and a positive correlation in the low (Figure 5D: $r_{(22)} = 0.60$, $\beta = 0.12$, 95% *CI* [0.26, 0.81]) anxiety subset. Closer inspection of the data confirmed the difference between the two slopes, which suggests that the rate of change in rsFC with unit increment of BIS was higher in the low anxiety subset as compared to the high anxiety subset.

A similar moderating effect of anxiety was observed for Recall-linked Cognitive Control of neutral memories, wherein the association between rsFC and Recall-linked Cognitive Control weakened with increasing anxiety. Specifically, significant interactions were identified between the right inferior frontal gyrus pars triangularis (IFG r) seed and two clusters. Cluster 1 encompassing the left precentral gyrus and posterior cingulate cortex (PC; $T_{(40)} = -6.16$, *p*FWE = 0.003, cluster size *kE* = 346, peak at $x = -16$, $y = -26$, $z = +38$; Figure 6A) and cluster 2 encompassing pSMG right and middle temporal gyrus, temporooccipital part right (toMTG r; $T_{(40)} = -5.22$, *p*FWE = 0.015, cluster size *kE* = 261, peak at $x = +58$, $y = -44$, $z = +04$; Figure 6C). Another interaction was observed between right posterior supramarginal gyrus seed and bilateral superior frontal gyrus (SFG l/r; $T_{(40)} = -6.66$, *p*FWE = 0.008, cluster size *kE* = 304, peak at $x = -10$, $y = +32$, $z = +48$; Figure 6E).

Separate visualizations conducted separately within anxiety subsets for the two target clusters revealed distinct association patterns between Recall-linked Cognitive Control of neutral memories or BIS and rsFC. Cluster 1 showed a negative correlation in the high (Figure 6B: $r_{(21)} = -0.36$, $\beta = -0.06$, 95% *CI* [-0.67, 0.06]) and a positive correlation in the low (Figure 5B: $r_{(22)} = 0.66$, $\beta = 0.12$, 95% *CI* [0.34, 0.84]) anxiety subset. Similarly, cluster 2 exhibited a negative correlation in the high (Figure 6D: $r_{(21)} = -0.10$, $\beta = -0.01$, 95% *CI* [-0.49, -0.32]) and a positive correlation in the low (Figure 6D: $r_{(22)} = 0.61$, $\beta = 0.12$, 95% *CI* [0.27, 0.81]) anxiety subset. Another interaction effect observed in posterior supramarginal gyrus exhibited negative correlation in the high (Figure 6F: $r_{(21)} = -0.68$, $\beta = -0.11$, 95% *CI* [-0.85, -0.37]) and a positive correlation in the low (Figure 6F: $r_{(22)} = 0.64$, $\beta = 0.12$, 95% *CI* [0.31, 0.83]) anxiety subset.

Closer inspection of the data confirmed the difference between the two slopes, which suggests that the rate of change in rsFC with unit increment of BIS was higher in the low anxiety subset as compared to the high anxiety subset.

No significant associations were observed between Recall-linked Cognitive Control of negative memories and rsFC.

------------------------------------------------------------------------------------------------------------------------------------

Insert Figure 5 & 6

------------------------------------------------------------------------------------------------------------------------------------

### 3.2.5. ***Differential Resting-State Functional Connectivity Supporting Suppression Versus Recall-linked Cognitive Control of Emotional Memories***

For Suppression-linked Cognitive Control of positive memories, we observed significantly greater rsFC between the right hippocampal seed and a cluster encompassing the thalamus and caudate ($T_{(41)}$ = 5.26, *p*FWE = 0.009, cluster size *kE* = 289, peak at x = −14, y = −10, z = +16; Figure 7A). To illustrate the direction and magnitude of the effect, rsFC values extracted from this cluster showed a positive association with Suppression-linked Cognitive Control of positive memories or BIS (Figure 7B: $r_{(45)}$ = 0.43, $\beta$ =0.06, 95% *CI* [0.16, 0.64]) and a negative association with Recall-linked Cognitive Control of positive memories or BIS (Figure 7B: $r_{(45)}$ = -0.43, $\beta$ =-0.06, 95% *CI* [-0.64, -0.16]). These findings indicate that greater Suppression -linked Cognitive Control of positive memories was associated with increased hippocampal coupling with thalamic regions.

No significant patterns of differential connectivity between Suppression and Recall-linked Cognitive Control were observed for negative or neutral memories.

-----------------------------------------------------------------------------------------------------------------------------------

Insert Figure 7

-----------------------------------------------------------------------------------------------------------------------------------

## 4. Discussion

In this study, we investigated volitional control over emotionally valenced memories by contrasting directed Suppression and Recall, and by characterizing the large-scale rsFC networks that support these processes when challenged by a concurrent visual working memory demand. Our findings reveal dissociable rsFC profiles associated with Suppression and Recall-linked Cognitive Control that varied as a function of emotional valence and were systematically modulated by individual differences in subclinical anxiety.

More efficient Suppression of positive memories was associated with reduced rsFC between an anterior cingulate cortex seed within the salience network and posterior regions encompassing the intracalcarine cortex, precuneus, and supracalcarine cortex, as well as diminished connectivity between the hippocampus and frontal pole. These patterns may aid in attenuation of perceptual, imagery-related, and mnemonic reinstatement processes during greater Suppression under a concurrent visual WM load. In contrast, greater cognitive control in suppressing negative memories was positively associated with connectivity between the posterior supramarginal gyrus and a cluster encompassing the angular gyrus and superior lateral occipital cortex. It could be argued that this brain-behaviour correlation pattern may be linked with greater reliance on attentional reorienting and perceptual control processes.

Anxiety significantly moderated several network–behaviour relationships. Higher anxiety was associated with increased coupling between superior frontal and medial frontal cortices as Suppression-linked Cognitive Control for positive memories increased, consistent with enhanced or compensatory engagement of prefrontal control systems. Anxiety also moderated Recall-related connectivity for positive and neutral memories, such that greater Recall-linked Cognitive Control was accompanied by reduced coupling between inferior frontal gyrus pars triangularis and distributed frontal, temporal, and parietal regions implicated in salience detection, inhibition, and emotion regulation. Moderation effects involving Suppression of negative memories and other Recall contrasts did not survive correction. Direct comparison of Suppression and Recall-linked Cognitive Control revealed increased hippocampal–thalamic connectivity during better Suppression of positive memories, that may be associated with heightened control demands on memory regulation.

Collectively, these findings extend prior task-based neuroimaging work by demonstrating that functional brain network organization associates with volitional regulation of emotional memories under competing cognitive demands and that anxiety selectively shapes these network-level functional patterns.

While the present findings highlight large-scale networks linked to emotional memory control and their selective modulation by anxiety, an important consideration in their interpretation is the gender imbalance in the sample (male:female ≈ 3:1; see Table 1), particularly given established sex differences in anxiety and emotional processing. Prior work indicates that females may exhibit differences in emotional reactivity and large-scale functional organization, including intrinsic connectivity within default mode regions such as the precuneus and posterior cingulate cortex, as well as engagement of prefrontal regulatory systems (Sacher et al., 2013; Stevens & Hamann, 2012). In this context, the functional networks identified in the present study, including anterior cingulate-posterior midline interactions during Suppression, hippocampal-frontal and hippocampal-thalamic coupling, and anxiety-modulated prefrontal connectivity involving superior and inferior frontal regions overlap with systems implicated in affective and self-referential processing. Particularly, the observed anxiety-related modulation of superior and inferior frontal connectivity during Suppression and Recall is consistent with evidence for sex differences in prefrontal-limbic engagement (Stevens & Hamann, 2012) , while the involvement of precuneus and posterior cingulate regions corresponds to reports of stronger default mode mode network connectivity in females (Bluhm et al., 2008; Sacher et al., 2013). We included gender as a nuisance covariate in the rsFC analyses which does not allow inferences regarding sex-specific neural mechanisms. Accordingly, the present findings should be interpreted as reflecting general patterns within the sampled population, with the possibility that the magnitude or direction of these network-level associations may differ in more gender-balanced samples. Future work specifically designed to examine sex differences will be essential to determine whether anxiety-related modulation of these networks during emotional memory control is differentially expressed across females and males.

### **4.1. *Resting-State Functional Networks Supporting Directed Suppression-linked Cognitive Control of Emotional Memories Under* a *concurrent visual WM load***

Our findings indicate that directed Suppression of emotional memories engages valence-sensitive recalibration of functional brain networks that support cognitive control under working memory load. Better cognitive control during Suppression of positive memories was associated with altered rsFC between the anterior cingulate cortex (ACC), a key node of the salience network implicated in cognitive control and emotion regulation (Botvinick et al., 2001; Crespo-García et al., 2022; Rolls, 2019), and posterior regions including the intracalcarine cortex, precuneus, and supracalcarine cortex. These posterior regions contribute to visual imagery, episodic retrieval, and internally generated representations (Cavanna & Trimble, 2006; Waldhauser et al., 2016). The negative shift in ACC-occipital coupling with increasing Suppression-linked Cognitive Control may be associated with reduced engagement of perceptual and imagery-related processes. This pattern could potentially relate to reduced mnemonic reinstatement and diminishing interference from emotionally salient content under external load.

Suppression-linked Cognitive Control of positive memories also associated with reduced connectivity between the hippocampus and frontal pole. The frontal pole supports cognitive branching and the management of internally generated mental states without disrupting ongoing task demands (Tsujimoto et al., 2011). Through its connections with anterior temporal regions and the hippocampus (Liu et al., 2013; Ramnani & Owen, 2004), the frontal pole may contribute to regulating hippocampal retrieval processes during Suppression. This regulation may involve GABA-mediated inhibitory mechanisms within the hippocampus (Schmitz et al., 2017). While dorsolateral prefrontal–hippocampal interactions in memory inhibition are well established (Anderson & Hanslmayr, 2014; Gagnepain et al., 2017), the present findings extend this framework by providing support that the frontal pole may also play a role in Suppression-related control, particularly when memory regulation requires coordination with concurrent cognitive demands.

By contrast, greater cognitive control during Suppression of negative memories was characterized by increased connectivity between the posterior supramarginal gyrus and regions encompassing the angular gyrus and superior lateral occipital cortex. The supramarginal gyrus is implicated in attentional reorientation away from internally generated content toward task-relevant stimuli (Cabeza et al., 2008; Carter & Huettel, 2013) and in emotion regulation (Wada et al., 2021). The angular gyrus contributes to episodic retrieval and conflict processing (Seghier, 2013), whereas lateral occipital regions support visual processing. Together, these findings suggest that ssuppression of negative memories may rely more strongly on attentional and perceptual control processes, reflecting valence-dependent strategies for limiting emotional interference under load.

### 4.2. *Anxiety-Modulated Resting-State Networks Supporting Directed Suppression and Recall-linked Cognitive Control of Emotional Memories Under concurrent visual WM load*

Anxiety significantly moderated the relationship between Suppression-linked Cognitive Control for positive memories and rsFC. Individuals with higher anxiety exhibited increased coupling between the superior frontal gyrus, implicated in inhibitory control and higher-order regulation, and medial frontal cortex regions involved in emotion regulation. Although altered connectivity within these regions has been reported across psychopathology, our findings suggest a context-dependent pattern whereby anxious individuals exhibit enhanced prefrontal engagement during Suppression of positive memories. This interpretation is consistent with evidence that anxiety is associated with diminished vividness and accessibility of positive memories (Moscovitch et al., 2011) and extends prevailing accounts that have predominantly emphasized exaggerated processing of negative information.

Directed Recall also exhibited distinct anxiety-related modulation. During Recall of positive memories, greater cognitive control in individuals with higher anxiety associated with reduced connectivity between the inferior frontal gyrus pars triangularis, which directs attention toward salient information (Hu & Dolcos, 2017), and a distributed network including the pars opercularis (inhibitory control; Boen et al., 2022), middle temporal gyrus (memory retrieval; Lu et al., 2022), and anterior supramarginal gyrus (emotion regulation; Wada et al., 2021). This pattern suggests lower engagement of networks typically linked with elaboration and salience of positive memories, potentially corresponding to less vivid representations and reduced interference with concurrent task demands.

Similarly, during Recall of neutral memories, higher anxiety was associated with reduced connectivity between inferior frontal regions and posterior midline and parietal areas, including the posterior cingulate cortex and supramarginal gyrus, which support contextual integration and emotional appraisal. These anxiety-related reductions in coupling may co-occur with impoverished sensory and contextual representations during retrieval, potentially contributing to the tendency for anxious individuals to interpret neutral stimuli as ambiguous or threatening (Park et al., 2016). By contrast, anxiety was not associated with connectivity differences within the predefined regions of interest, suggesting that anxiety-related differences in negative retrieval, if present, may involve alternative or more distributed circuits not captured by our analyses.

### 4.3. *Resting-State Functional Networks Underlying Cognitive Control linked to Directed Suppression Versus Recall of Emotional Memories*

Direct comparison of Suppression and Recall revealed greater hippocampal–thalamic connectivity associated with efficient Suppression of positive memories, besides a weaker trend toward reduced hippocampal–frontal pole coupling. Beyond its role as a sensory relay, the thalamus supports maintenance and updating of internally generated representations and mediates hippocampal–prefrontal communication (Wolff & Vann, 2019). Given its involvement in working memory processes (Guo et al., 2017), enhanced hippocampal–thalamic connectivity may be linked to greater control demands during Suppression-linked Cognitive Control, possibly corresponding to sustained regulation of hippocampal reactivation under a concurrent visual WM load.

### 4.4. *Absence of Behavioural Effects*

Despite robust network-level differences, we did not observe significant behavioural differences in dual-task memory-linked interference / Cognitive Control between directed Suppression and Recall across emotional valences. This finding contrasts with prior evidence indicating that directed suppression can impair performance on a concurrent attentional task (Fawcett & Taylor, 2008). Several methodological differences may account for this discrepancy. Notably, we imposed a working memory load rather than an attentional load, and the concurrent task was not framed as secondary or subordinate in importance, potentially altering how cognitive resources were allocated between memory control and task performance. In addition, the absence of behavioural differences may reflect substantial interindividual variability in working memory capacity, which is known to influence the ability to manage intrusive thoughts. It is also possible that the working memory task employed here did not impose sufficient demands to elicit measurable differences in performance within the tested sample. Furthermore, we did not observe a significant association between anxiety and dual-task memory-linked interference / Cognitive Control (indexed by BIS scores), across emotional conditions. One possible explanation is that the BIS in the current design captured performance under concurrent task demands, introducing greater variability relative to paradigms that assess memory control in isolation and have previously demonstrated anxiety-related modulation of task efficiency (Dieler et al., 2014). Importantly, this study measured the influence of emotional memory control under competing demands in healthy (subclinical) participants, with relatively weak and short-lived emotional stimuli, who may likely have a higher threshold for responding to such subtle (less aversive / intense) stimuli.

### 4.5. *Limitations and future directions*

We identify a few limitations of the present study that warrant consideration. First, we did not model time of day as a nuisance covariate, despite evidence that circadian rhythms can influence rsFC. Second, the sample comprised a higher proportion of male participants (~3:1), which warrants caution in generalizing the findings, particularly given established gender differences in anxiety, emotional memory processing, and associated neural circuitry. Although gender was included as a covariate, this does not substitute for gender-balanced samples needed to determine whether the observed rsFC patterns show gender-specific modulation. Third, resting-state fMRI data were acquired at a single time point due to resource constraints and future work would benefit from repeated measurements across multiple sessions to establish the stability and reliability of the observed connectivity patterns. Fourth, although the sample size was statistically sufficient for the present analyses, it remains relatively modest, and replication in larger, independent samples will be essential to confirm the robustness of these effects. Finally, as rsFC is inherently correlational, the present findings do not permit causal inferences. Future studies employing perturbation-based approaches, such as non-invasive brain stimulation or pharmacological manipulations, may help to test the causal involvement of the identified networks and enhance its translational relevance.

## 5. *Conclusion*

Despite these limitations, the present study provides a relevant contribution by integrating resting-state functional connectivity with behavioural indices of volitional memory control under concurrent working memory load, while explicitly accounting for interindividual variability in subclinical anxiety. By situating emotional memory regulation within a network-level and cognitive resource-sharing framework, these findings extend existing task-based fMRI accounts and offer a complementary perspective on the intrinsic neural architecture associated with cognitive control of emotional memories.

### Data availability statement

All data that led to the reported results can be accessed here: https://doi.org/10.5281/zenodo.19443262

### Author Contributions:

**Shruti Kinger:** Conceptualization, Data Curation, Formal Analysis, Investigation, Methodology, Project Administration, Software, Visualization, Writing - Original Draft Preparation

**Mrinmoy Chakrabarty:** Conceptualization, Formal Analysis, Funding Acquisition, Investigation, Methodology, Project Administration, Software, Supervision, Validation, Visualization, Writing - Original Draft Preparation, Writing - Review & Editing

**Competing interests**

The authors declare no competing interests.

**Acknowledgments**

The authors thank all participants for their time and participation in the experiment. The authors are grateful to Manasi Chaturvedi (University of Texas, Austin, USA) and Dr. Suhail Rafiq Mir (All India Institute of Medical Sciences Delhi, INDIA) for their assistance and cooperation during data collection. The authors thank the Centre for Advanced Research in Imaging, Neuroscience and Genomics (CARING) of Mahajan Imaging, SDA, New Delhi, Delhi 110016, for providing the MRI imaging facility and Ms. Madhuri Barnwal, Mr. Baby, and Mr. Sibin for scheduling and technical assistance during scanning.

**Funding Information**

The research was funded by the intramural research Professional Development Allowance of IIIT-Delhi and the Science and Engineering Research Board Core Research Grant to Dr. Mrinmoy Chakrabarty (SERB-CRG/2022/008119). Ms. Shruti Kinger was supported by the institute PhD fellowship by IIIT-Delhi. The funders were not involved in the study design, data collection, analysis, interpretation, the writing of this article, or the decision to submit it for publication.

**Figures are attached below -**

**Figure 1**

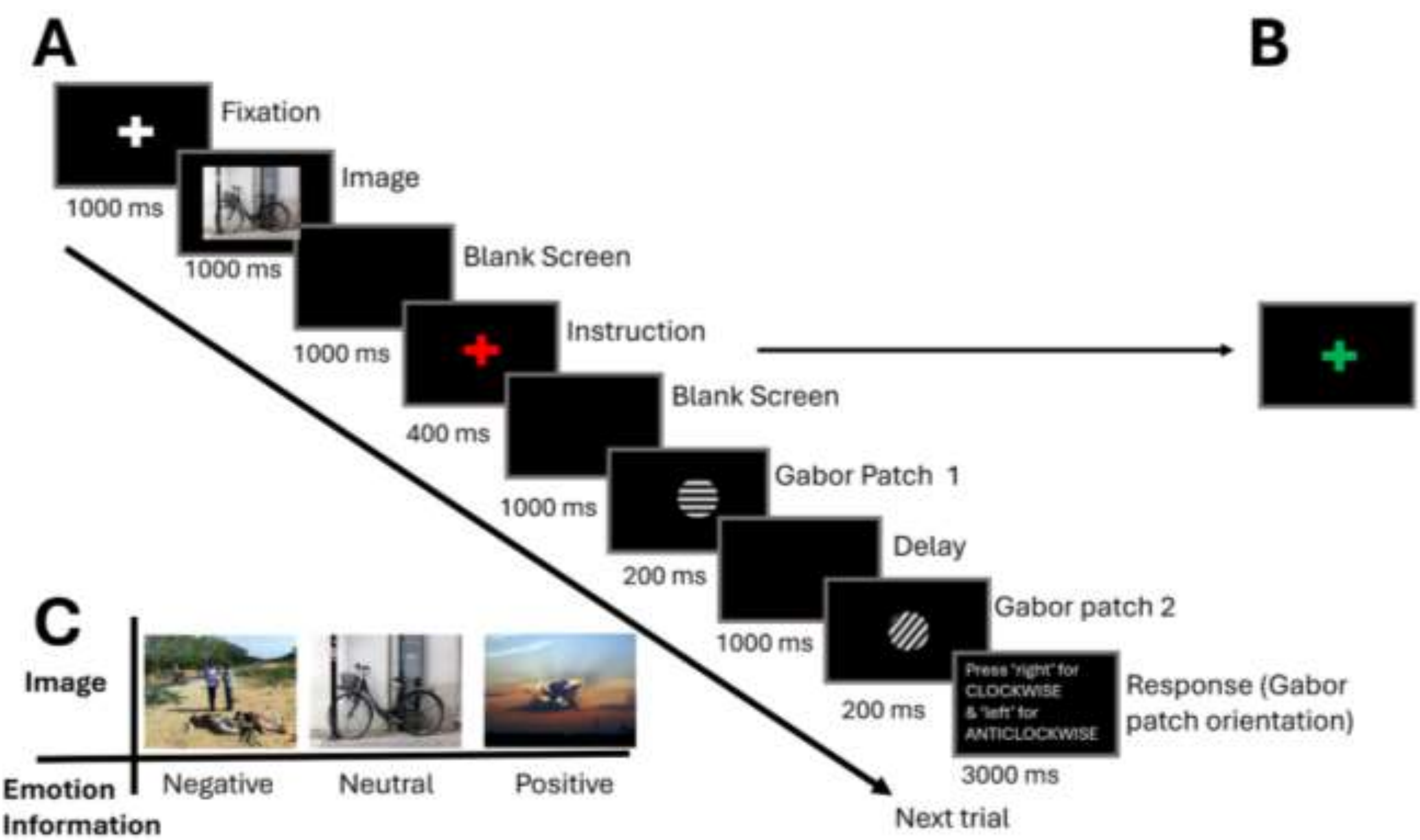

**Schematic of the behavioural task design.** The task began with a fixation point presented for 1000 ms followed by an image for another 1000 ms. A blank period followed for 1000 ms after which the fixation point appeared for 400 ms either in red colour (A) indicating 'No-think'/'Suppress' condition or green colour (B) indicating 'Think'/'Recall' condition. After another blank screen of 1000 ms, the independent visual working memory task was presented. Two gabor patches were presented each for 200 ms with a delay of 1000 ms. During the response window of 3000 ms, participants were asked to respond 'right' if the second gabor patch tilted clockwise relative to the first one and 'left' if the tilt was anticlockwise. (C) Three different exemplar images sourced from NAPS database conveying three emotion signals.

**Figure 2**

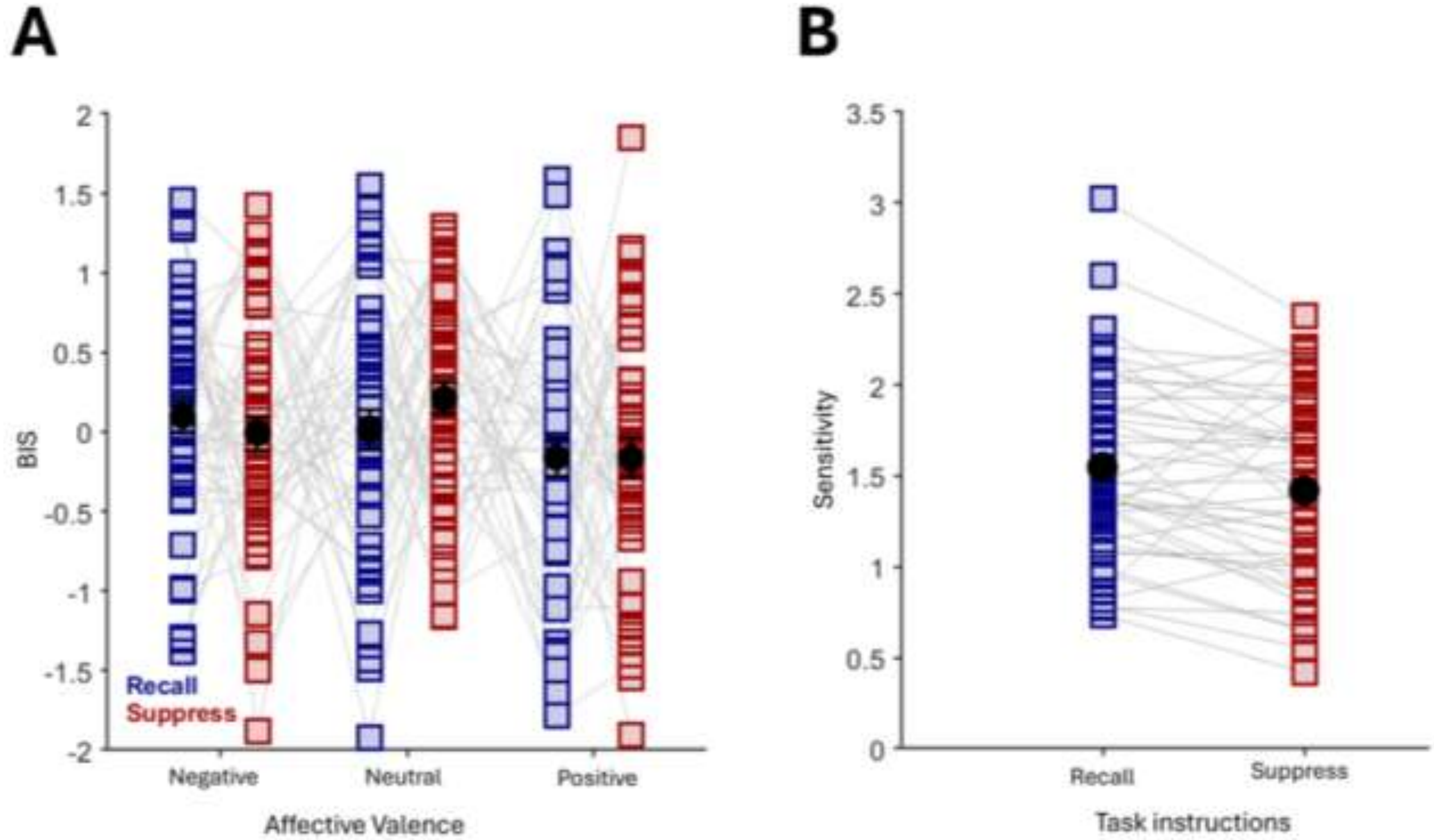

**Effects of task instructions (Recall and Suppress) and/or emotions (negative, neutral, and positive) on dual-task memory-linked interference or Cognitive Control / sensitivity**. (A) Plot showing dual-task memory-linked interference or Cognitive Control (BIS) of each participant linked with Recall condition and Suppress conditions in blue and red square respectively and joined with grey lines, mean across participants in black circle, and standard error of the mean as error bars in black. (B) Plot depicts the sensitivity measure of each participant for Recall (blue square) and Suppress (red square) instructions, mean across participants in black circles, and standard error of the mean as error bars in black.

**Figure 3**

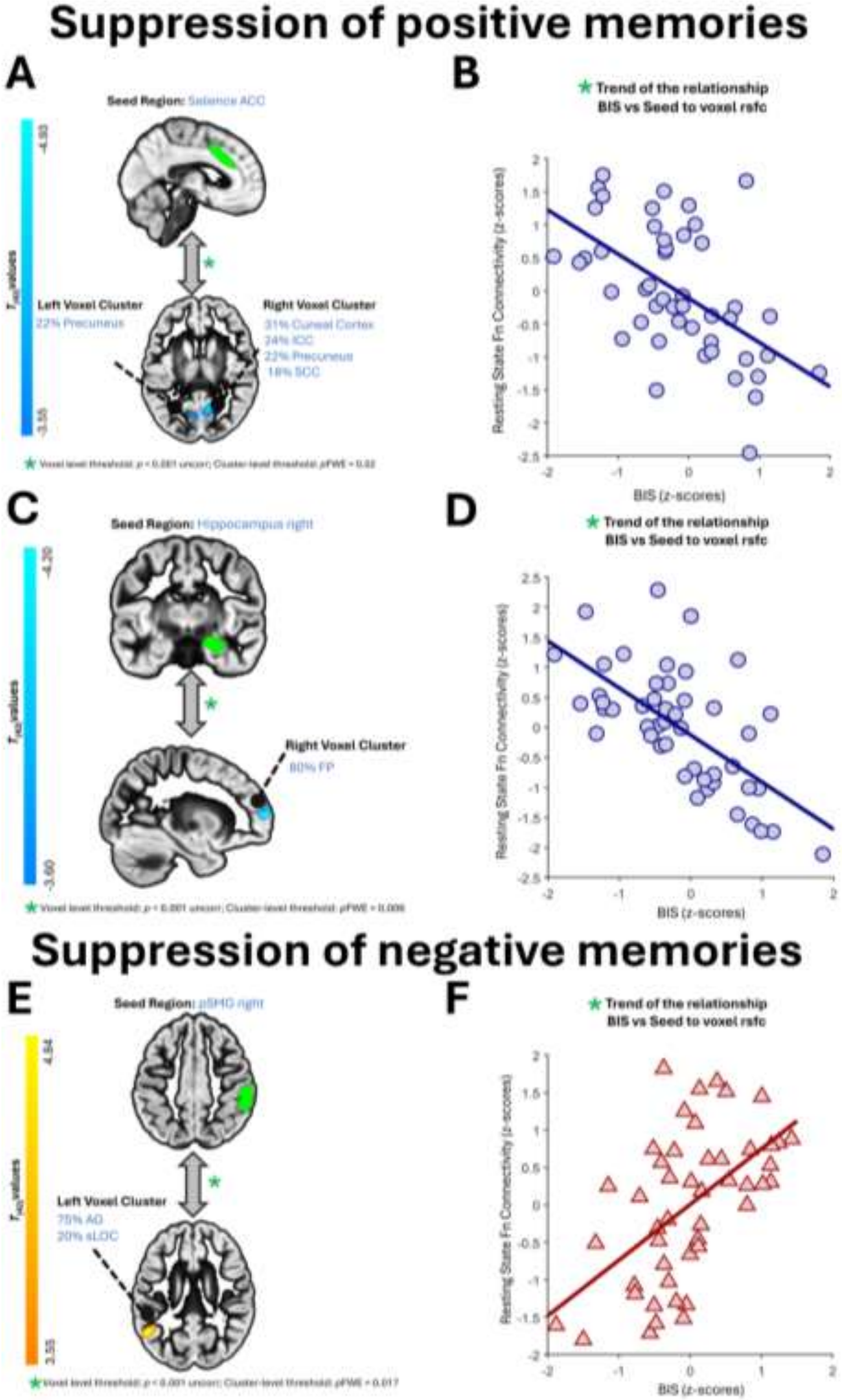

**Association between Suppression-linked Cognitive Control or BIS and rsFC.** (A, C, E) rsFC between seed region and the voxels contained in the cluster. The colourbars depict the direction and strength connectivity between seeds and the respective clusters. The cooler colourbars (A, C) depict negative correlation and warmer colourbar (E) depict positive correlation. (B, D, F) Scatter plots of the associations between BIS and rsFC shown in A, C, E respectively. Each marker depicts one participant, and least square lines show the trend in the associations. ICC: Intracalcarine Cortex, SCC: Supracalcarine Cortex, FP: Frontal Pole, pSMG: posterior division Supramarginal gyrus, AG: Angular Gyrus, sLOC: superior division of the lateral occipital cortex. *** *p*FWE < 0.05**

**Figure 4**

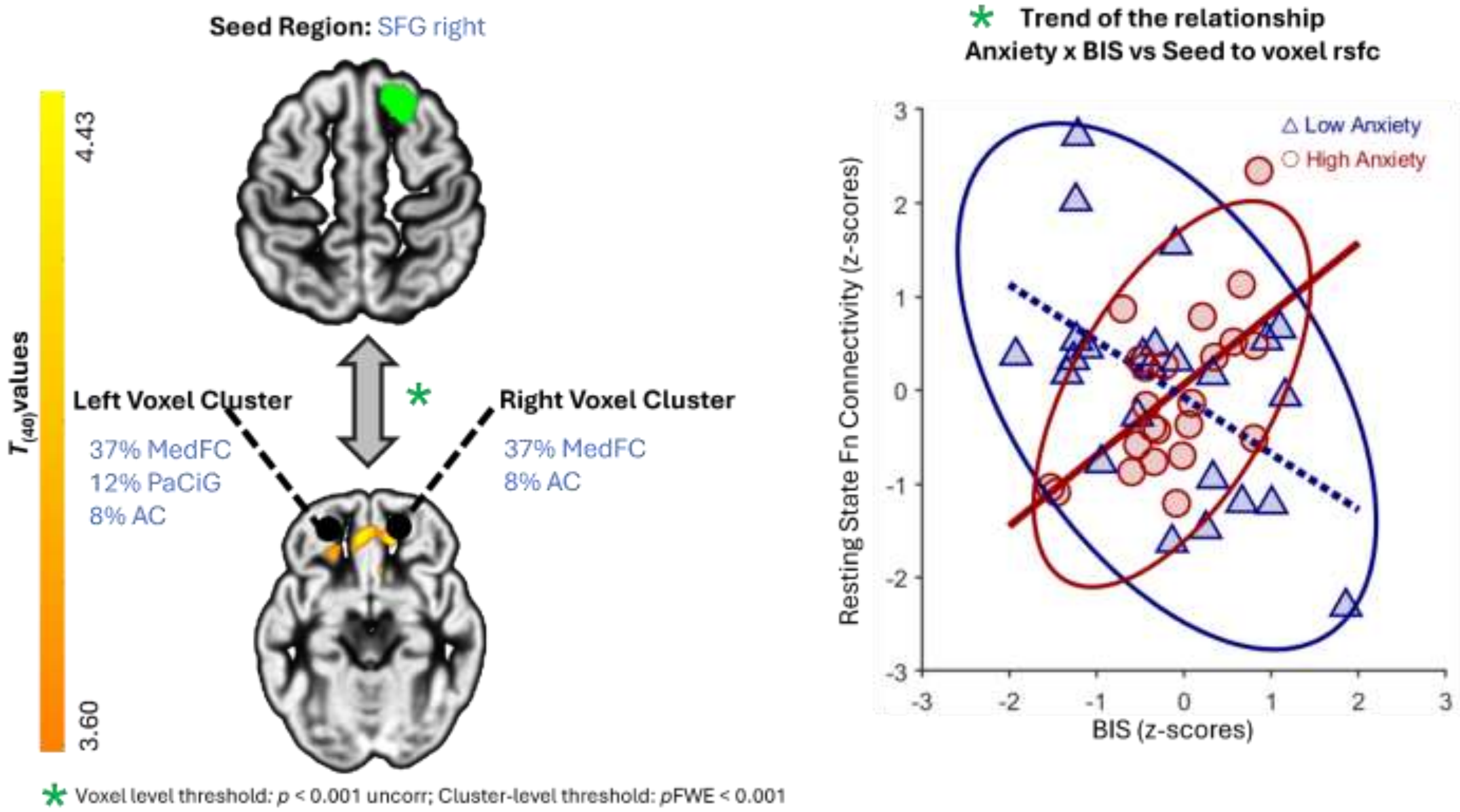

**Moderating effect of anxiety on the association between Suppression-linked Cognitive Control of positive memories or BIS and rsFC.** (A) rsFC between seed region and the cluster. The colourbar depicts the direction and strength of connectivity. The warmer colourbar depicts positive correlation. (B) Scatter plot of the associations between BIS and rsFC shown in A respectively. Each marker depicts one participant, and least square lines show the trend in the associations in high (red circle and solid line) and low (blue triangle and dotted line) anxiety subset. The confidence ellipses represent 95% confidence intervals and was plotted using the MATLAB function - *error_ellipse* (Johnson, 2026). SFG: Superior Frontal Gyrus, MedFC: Medial Frontal Cortex, PaCiG: Paracingulate Gyrus, AC: Anterior Cingulate. *** *p*FWE < 0.05**

**Figure 5**

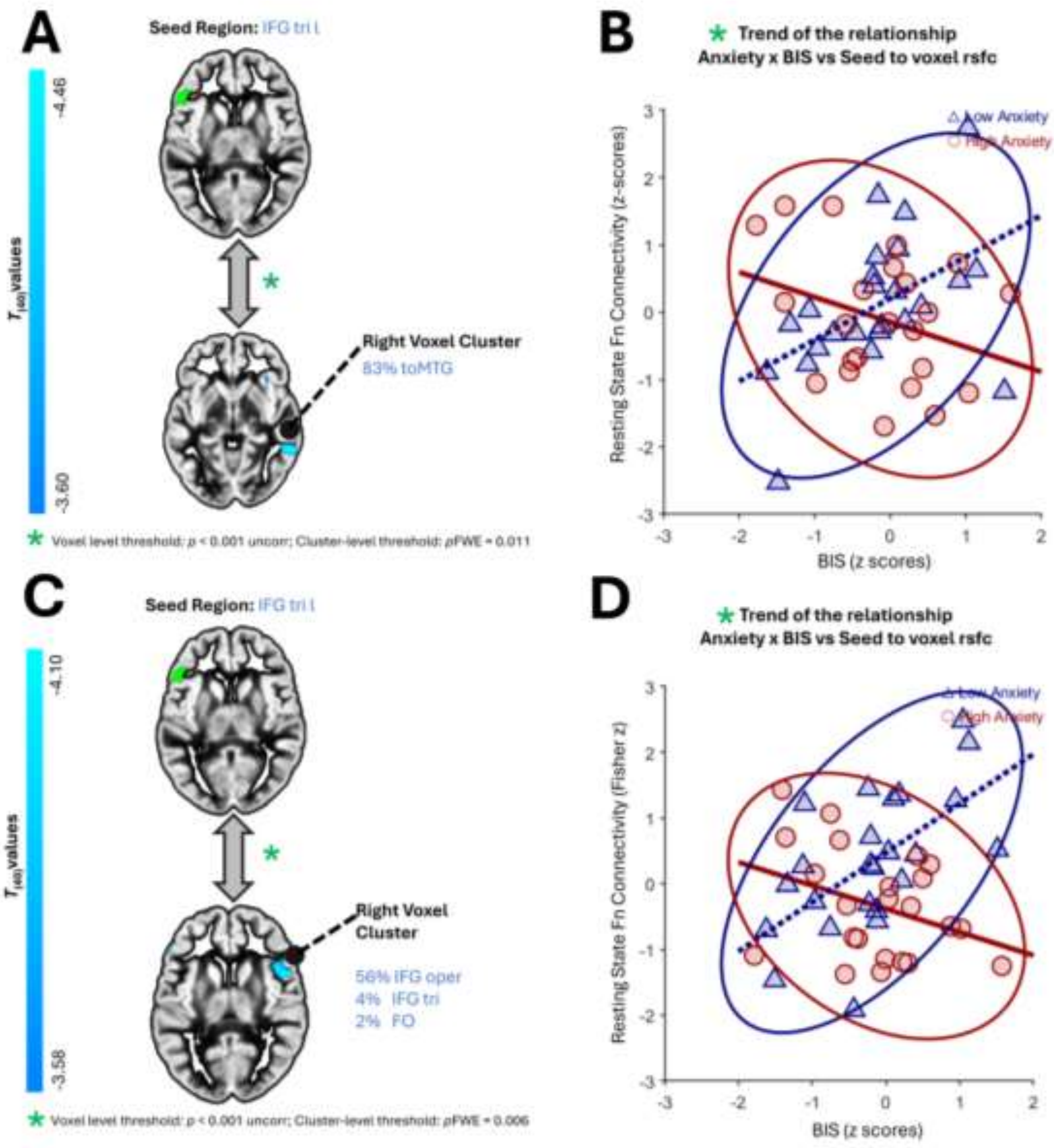

**Moderating effect of anxiety on the association between Recall-linked Cognitive Control of positive memories or BIS and rsFC.** (A, C, E) rsFC between seed region and the cluster. The colourbar depicts the direction and strength of connectivity. The cooler colourbars (A, C, E) depict negative correlation. (B, D, F) Scatter plots of the associations between BIS and rsFC shown in A, C, E respectively. Each marker depicts one participant, and least square lines show the trend in the associations in high (red circle and solid line) and low (blue triangle and dotted line) anxiety subset. The confidence ellipses represent 95% confidence intervals and was plotted using the MATLAB function - *error_ellipse* (Johnson, 2026). IFG oper: Inferior Frontal Gyrus opercularis, IFG tri: Inferior Frontal Gyrus triangularis, FO: Frontal Operculum, toMTG: temporal-occipital-middle temporal gyrus. *** *p*FWE < 0.05**

**Figure 6**

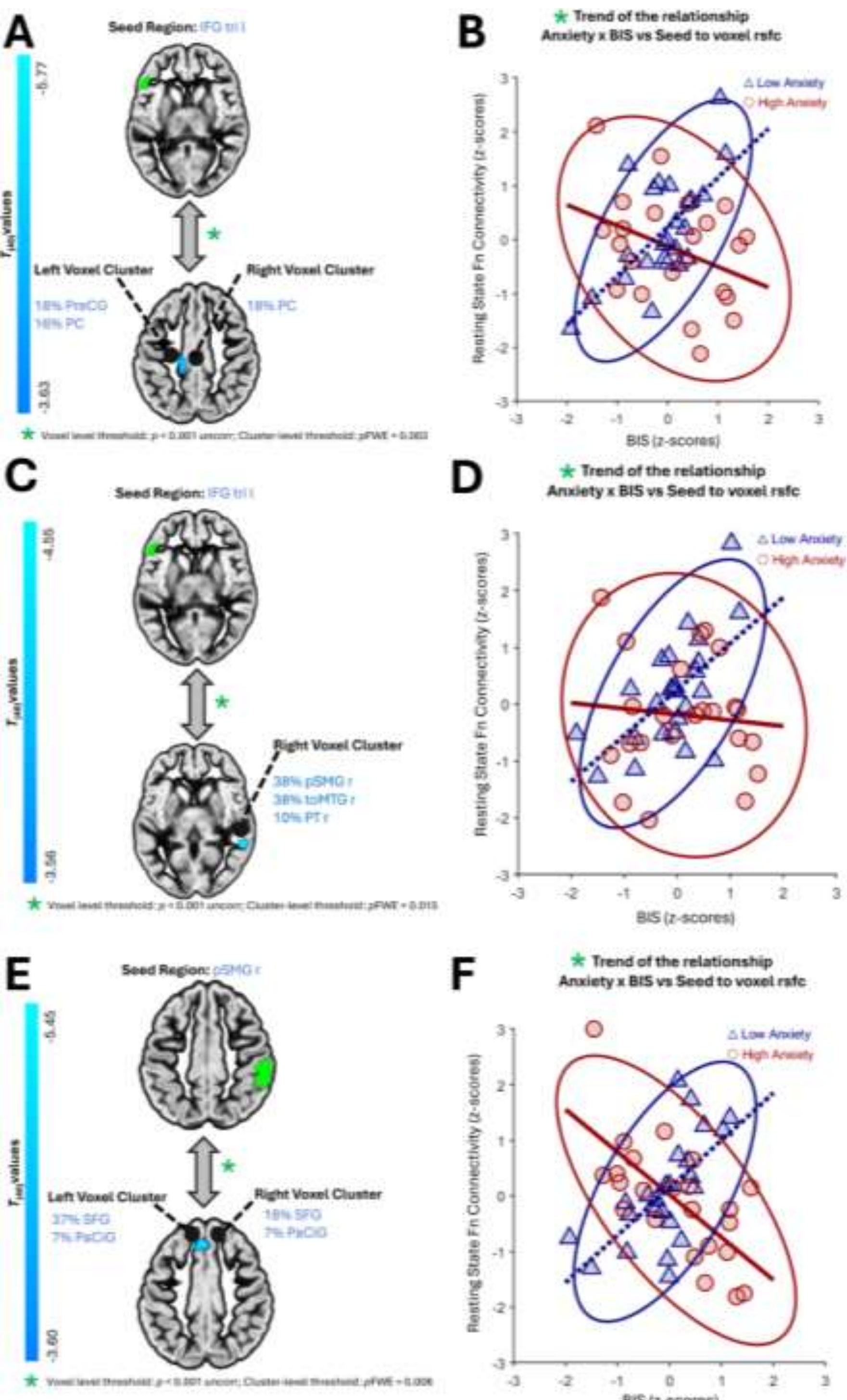

**Moderating effect of anxiety on the association between Recall-linked Cognitive Control of neutral memories or BIS and rsFC.** (A, C) rsFC between seed region and the cluster. The colourbar depicts the direction and strength of connectivity. The cooler colourbars (A, C) depict negative correlation. (B, D) Scatter plots of the associations between BIS and rsFC shown in A and C respectively. Each marker depicts one participant, and least square lines show the trend in the associations in high (red circle and solid line) and low (blue triangle and dotted line) anxiety subset. The confidence ellipses represent 95% confidence intervals and was plotted using the MATLAB function - *error_ellipse* (Johnson, 2026).... PreCG: Precentral Gyrus, PC: Posterior Cingulate, pSMG:

posterior Supramarginal Gurus, toMTG: temporo-occipital Middle Temporal Gyrus SFG: Superior Frontal Gyrus, PaCiG: Paracingulate Gyrus. *** *p*FWE < 0.05**

**Figure 7**

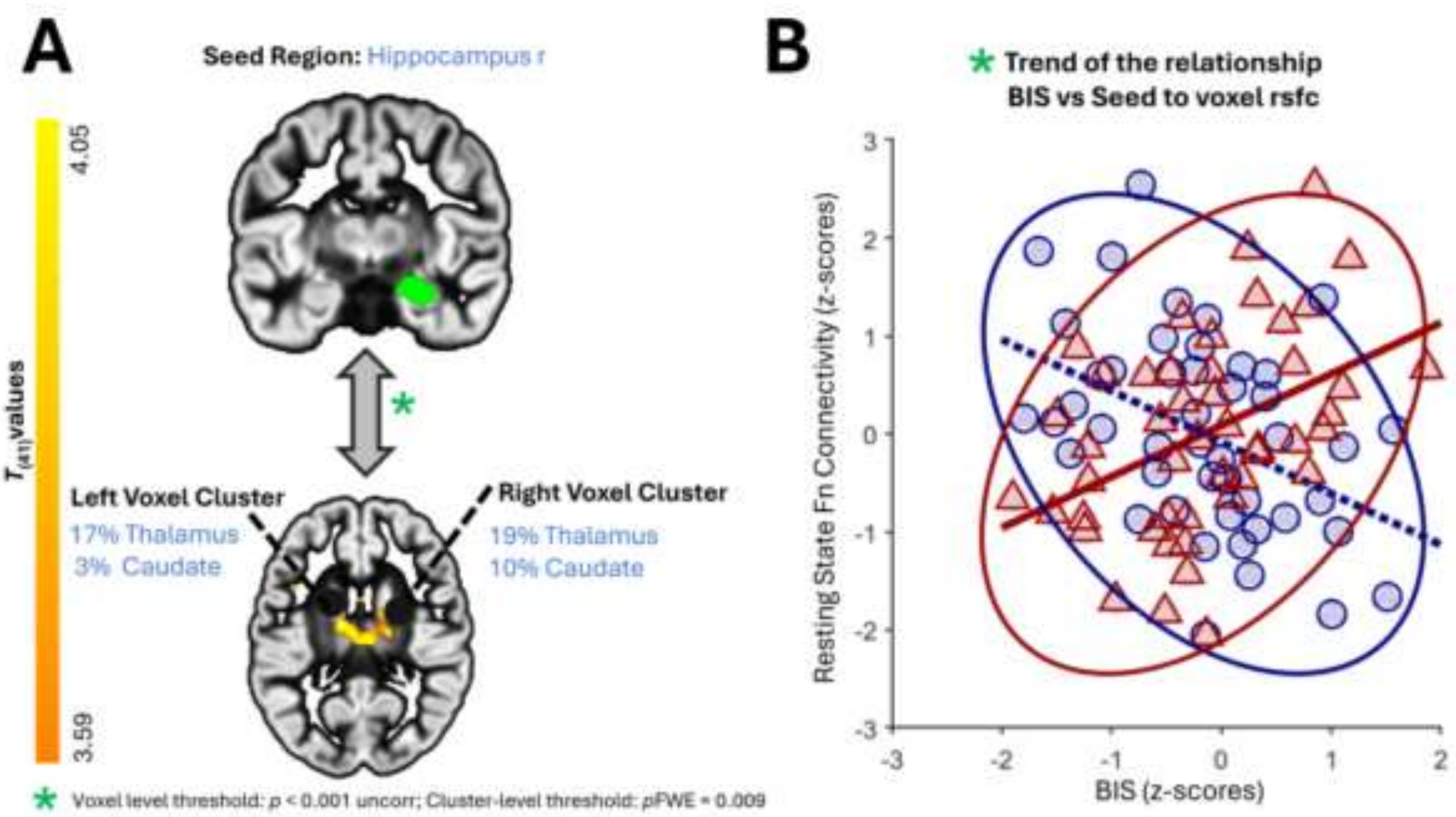

**Differential connectivity pattern observed for Suppression vs. Recall-linked Cognitive Control of positive memories or BIS.** (A) rsFC between seed region and the cluster. The colourbar depicts the direction and strength of connectivity. The warmer colourbar (A) depicts positive correlation. (B) Scatter plots of the associations between BIS and rsFC shown in A. Each marker depicts one participant, and least square lines show the trend in the associations for Suppression (red triangle and solid line) and Recall (blue circle and dotted line). The confidence ellipses represent 95% confidence intervals and was plotted using the MATLAB function - *error_ellipse* (Johnson, 2026). *** *p*FWE < 0.05**

**Tables are attached below**

**Table 1**

| Particulars | Ratio / Mean ± SD |
|---|---|
| Gender ratio (male: female) | ~3:1 (33:14) |
| Age (in years) | 21.66 ± 2.43 |
| PHQ | 7.17 ± 4.46 |
| Fear Affect | 14 ± 7 |

Demographic information of the participants. PHQ: Patient Health Questionnaire

**Table 2**

| S. No. | Hypotheses | Keywords | A priori ROIs |
|---|---|---|---|
| 1. | Suppression-linked Cognitive Control of emotionally charged memories is associated with resting state functional connectivity. (see Equation 2) | memory suppression, emotion, fMRI / neuroimaging / rsfMRI, cognitive load | hippocampus ($w$ = 0.21), anterior cingulate cortex ($w$ = 0.15), posterior supramarginal gyrus ($w$ = 0.11) |
| 2. | Recall-linked Cognitive Control of emotionally charged memories is associated with resting state functional connectivity. (see Equation 2) | memory retrieval, emotion, fMRI / neuroimaging / rsfMRI, cognitive load | amygdala ($w$ = 0.24), precuneus ($w$ = 0.23), hippocampus ($w$ = 0.09) |
| 3. | Anxiety moderates the association between Suppression-linked Cognitive Control of emotionally charged memories and resting state functional connectivity. (see Equation 3) | memory suppression, emotion, fMRI / neuroimaging / rsfMRI, cognitive load, anxiety | superior frontal gyrus ($w$ = 0.26), middle frontal gyrus ($w$ = 0.26), pars triangularis inferior frontal gyrus ($w$ = 0.26) |
| 4. | Anxiety moderates the association between Recall-linked Cognitive Control of emotionally charged memories and resting state functional connectivity. (see Equation 3) | memory retrieval, emotion, fMRI / neuroimaging / rsfMRI, cognitive load, anxiety | posterior supramarginal gyrus (w = 1), amygdala (w = 0.41), inferior frontal gyrus (w =0.19) |

| 5. | rsFC differs between Suppression and Recall-linked Cognitive Control. (see Equation 4) | memory retrieval, memory suppression, visual stimuli, emotion, fMRI / neuroimaging / rsfMRI, cognitive load | hippocampus (*w* = 0.20), middle frontal gyrus (*w* = 0.16), anterior cingulate cortex (*w* = 0.11) |
|---|---|---|---|

**MRI Hypotheses, Keywords, and corresponding NeuroQuery weights for A Priori ROIs:** The table summarizes the MRI hypotheses, the keywords used to screen ROIs selection on PubMed and the corresponding ROI-specific weights (w) for the queried cognitive constructs (entered as keywords) obtained using NeuroQuery. Please note that, as *memory suppression* cannot be queried as a single term in NeuroQuery, we used *forgetting* as a proxy to obtain the aforementioned weights.

**Table 3**

| Analysis | Seed | Clusters | MNI-coordinates<br><br>(X Y Z) | Cluster size (*kE*) | Cluster level *p*-FWE | Effect size (T-value) |
|---|---|---|---|---|---|---|
| Suppression -linked Cognitive Control of Positive memories vs rsfc regression | Salience ACC | +10 -66 +18 | | 246 | 0.02 | $T_{(42)} = -5.06$ |
| | | Cuneal r<br>Cuneal Cortex right | +14 -70 +22 | 76 | | |
| | | ICC r<br>Intracalcarine Cortex right | +10 -68 +14 | 58 | | |
| | | Precuneus | +10 -64 +18 | 55 | | |
| | | SCC r<br>Supracalcarine Cortex right | +12 -64 +16 | 18 | | |
| | Hippocampus right | +22 +56 +10 | | 317 | 0.006 | $T_{(42)} = -5.69$ |
| | | FP r<br>Frontal Pole right | +22 +64 +12 | 255 | | |
| Suppression -linked Cognitive Control of Negative memories vs rsfc regression | Supramarginal Gyrus posterior division right (pSMG r) | -52 -58 +22 | | 261 | 0.017 | $T_{(42)} = 4.64$ |
| | | AG l Angular Gyrus left | -52 -56 +24 | 197 | | |
| | | sLOC l<br>Lateral Occipital Cortex, superior division left | -54 -62 +24 | 52 | | |
| Suppression -linked Cognitive Control of Positive memories X | SFG r<br>Superior Frontal Gyrus right | -04 +44 -12 | | 617 | < 0.001 | $T_{(40)} = 5.81$ |
| | | MedFC Frontal Medial Cortex | +0 +44 -16 | 231 | | |
| anxiety vs rsfc regression | | PaCiG l Paracingulate Gyrus left | -6 +40 -8 | 73 | | |
| | | AC Cingulate Gyrus, anterior division | -4 +34 -4 | 50 | | |
| | | SubCalC (Subcallosal Cortex) | +6 +24 -10 | 27 | | |
| | | not-labelled | +8 +32 -8 | 233 | | |
| Recall-linked Cognitive Control of Positive memories X anxiety vs rsfc regression | IFG tri l Inferior Frontal Gyrus pars triangularis left | Cluster 1: +66 -50 -10 | | 296 | 0.008 | $T_{(40)} = -5.91$ |
| | | toMTG r Middle Temporal Gyrus, temporooccipit-al Right | +60 -46 -4 | 246 | | |
| | | Cluster 2: +46 +20 +18 | | 270 | | $T_{(40)} = -5.59$ |
| | | IFG oper r Inferior Frontal Gyrus, pars opercularis right | +50 +14 +14 | 151 | 0.012 | |
| | | IFG tri r Inferior Frontal Gyrus, pars triangularis right | +48 +22 +10 | 12 | | |
| | | FO r Frontal Operculum right | +44 +18 +8 | 5 | | |
| | | not-labelled | +44 +18 +16 | 101 | | |
| Recall-linked Cognitive Control of Neutral memories X anxiety vs rsfc regression | IFG tri l Inferior Frontal Gyrus pars triangularis left | Cluster 1: -16 -26 +38 | | 346 | 0.003 | $T_{(40)} = -6.16$ |
| | | PreCG l Precentral Gyrus left | -14 -30 +44 | 61 | | |
| | | PC Cingulate Gyrus, posterior division | -12 -28 +40 | 54 | | |
| | | not- labelled | -16 -28 +40 | 213 | | |
| | | Cluster 2: +58 -44 +04 | | 261 | 0.015 | $T_{(40)} = -5.22$ |
| | | pSMG r Supramarginal Gyrus, posterior division Right | +58 -38 +18 | 99 | | |
| | | toMTG r Middle Temporal Gyrus, temporooccipital part Right | +58 -44 +4 | 98 | | |
| | pSMG r Supramarginal Gyrus posterior division right | -10 +32 +48 | | 304 | 0.008 | $T_{(40)} = -6.66$ |
| | | SFG l Superior Frontal Gyrus left | -6 +32 +46 | 113 | | |
| | | SFG r Superior Frontal Gyrus right | +4 +32 +46 | 54 | | |
| | | PaCiG r Paracingulate Gyrus right | +4 +30 +44 | 22 | | |
| | | PaCiG l Paracingulate Gyrus left | -4 +32 +40 | 20 | | |
| | | not-labelled | -8 +32 +46 | 95 | | |

| Suppression -linked Cognitve Control of positive memories versus Recall of positive memories | Hippocampus right | Cluster 1:<br>-14 -10 +16 | | 289 | 0.009 | $T_{(41)}$ = 5.26 |
|---|---|---|---|---|---|---|
| | | Thalamus right | +10 -12 +18 | 56 | | |
| | | Thalamus left | -8 -14 +16 | 48 | | |
| | | Caudate right | +12 -2 +16 | 29 | | |
| | | Caudate left | -12 -8 +18 | 8 | | |
| | | not-labelled | +4 -12 +18 | 148 | | |

**MRI results:** Apriori regions that survived multiple correction after testing of MRI hypotheses provided in Table 2 ($p$FWE ≤ 0.017).

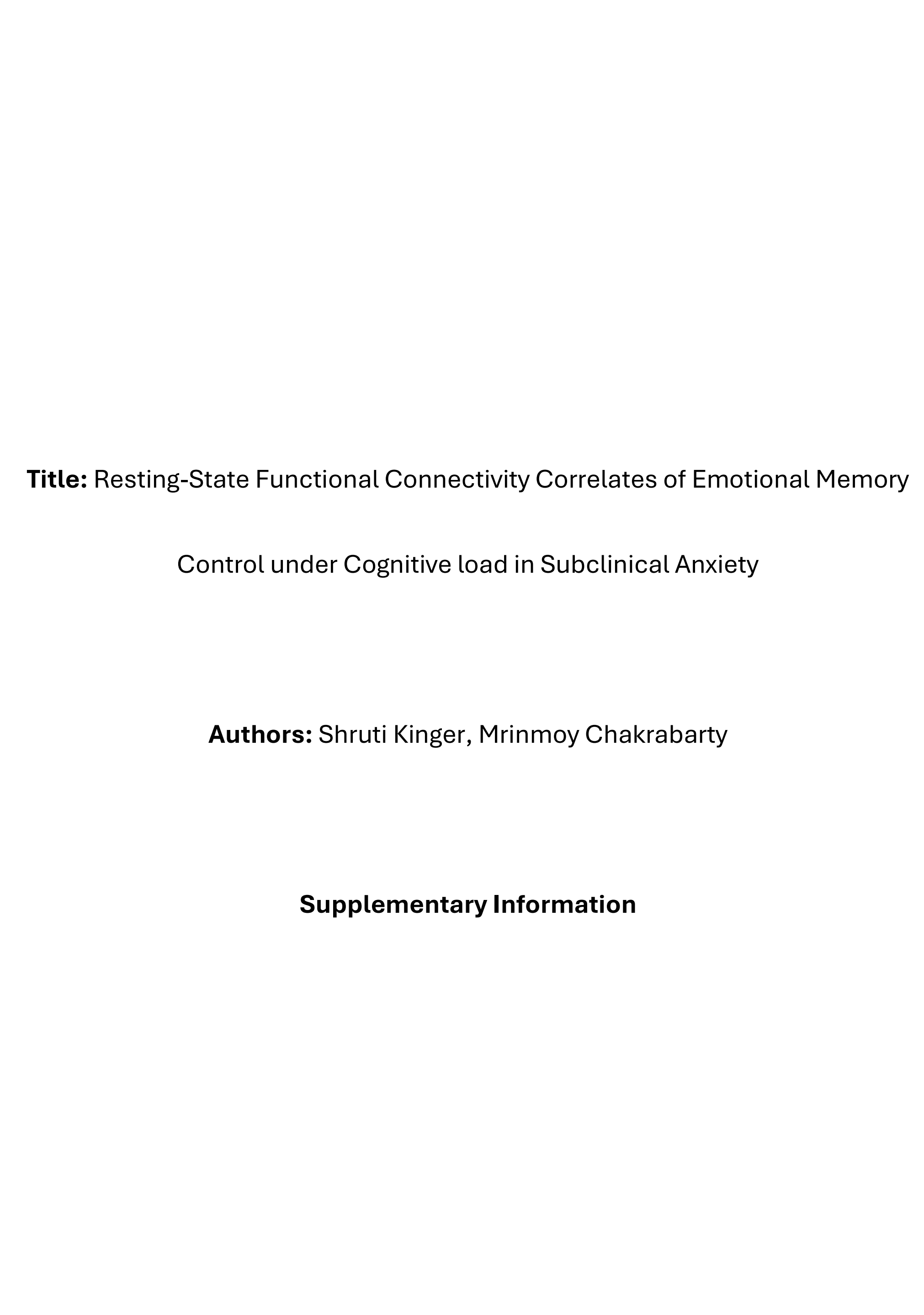

**Title:** Resting-State Functional Connectivity Correlates of Emotional Memory Control under Cognitive load in Subclinical Anxiety

**Authors:** Shruti Kinger, Mrinmoy Chakrabarty

## Supplementary Information

**S1 – Construct validity and Reliability of the Balance Integration Scores (BIS) metric**

**A) Convergent validity:** To ascertain whether the BIS metric correlates with a measure that also reflects interference, we analysed its relationship with the working memory capacity.

Working memory capacity is the amount of information that can be held and manipulated in immediate memory at the same time. It is also linked to interference such that higher working memory capacity indicates lower interference. We quantified working memory capacity (Cowan's *K*), defined as the number of task-relevant items held in and retrieved from working memory, across six conditions (Negative, Neutral, and Positive Recall; Negative, Neutral, and Positive Suppression) to examine its association with BIS. To this end, we kept the set size as 2 as the orientation of both the Gabor patches was task relevant. Subsequently, hit rate (H) was calculated as the proportion of correctly identified trials out of the total number of trials, whereas false alarm rate (FA) was calculated as the proportion of incorrect responses out of the total number of trials. Finally, we calculated Cowan's *K* using Equation 1.

$$K = N \times (H - FA) \ldots \text{(Equation 1)}$$

We observed a positive correlation between working memory capacity and BIS across all six conditions (all $r >= 0.24$). A higher working memory capacity in the directed Recall condition reflects weaker memory representations due to less successful memory retrieval and hence lower interference. In contrast, a higher working memory capacity in the directed Suppression condition indicates fewer intrusions due to more effective suppression and thus, lower interference. (Figure S1).

**Figure S1**

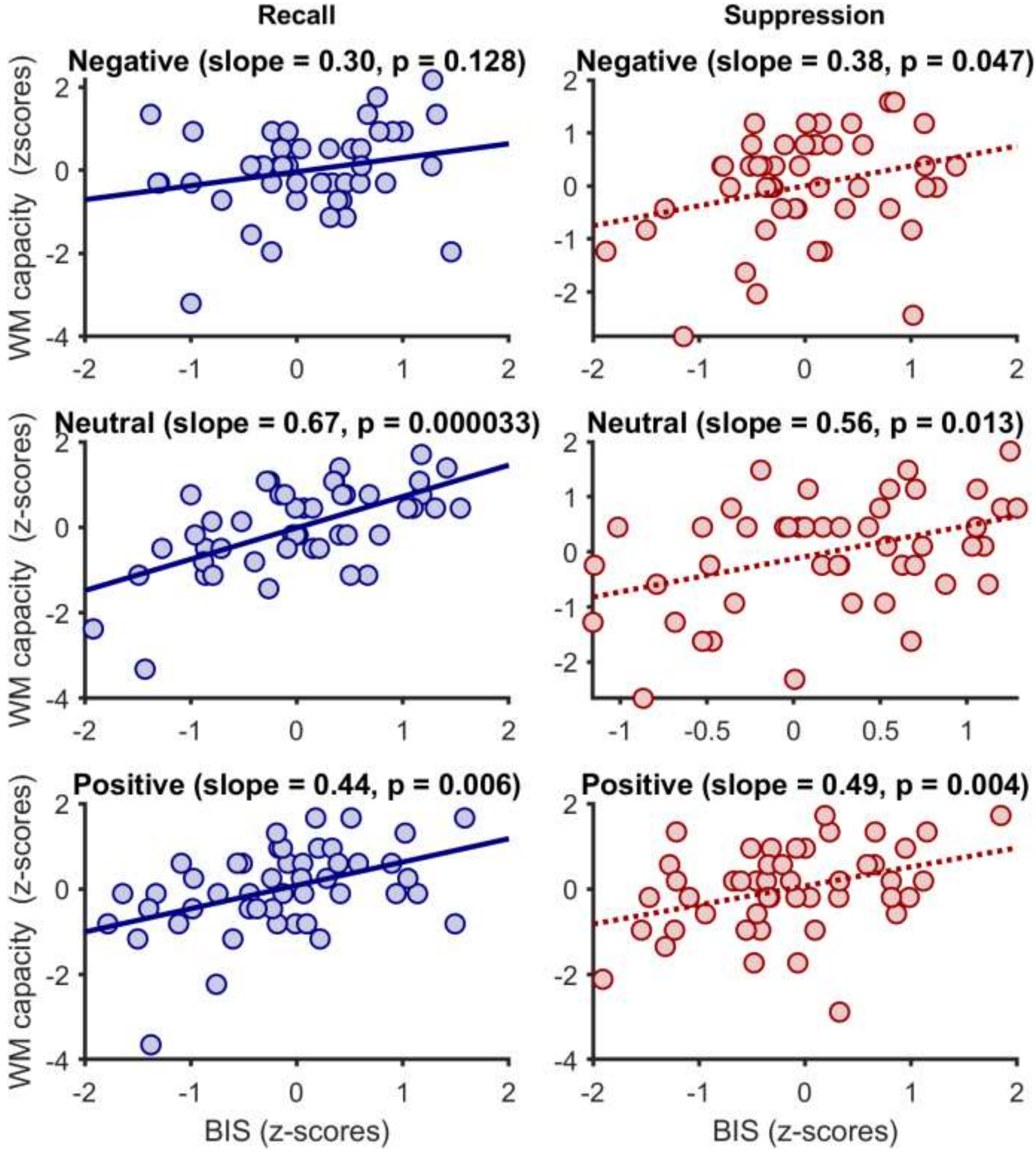

**Association between BIS and working memory (WM) capacity.** The left panels (A) show scatter plots between BIS and WM capacity (blue markers and solid fitted least square lines) for the Recall condition, while the right panels (B) show scatter plots between BIS and WM capacity (red markers and broken fitted least squares lines) for the Suppression condition. Each marker represents one participant. The slope and *p*-values were computed using the MATLAB function – *robustfit*.

**B) Discriminant validity:** To ascertain whether the BIS metric is unrelated to a measure that does not reflect interference, we analyzed its relationship with the sensitivity scores derived from the recognition task.

Sensitivity scores were computed based on participants' ability to correctly implement Recall and Suppression instructions and subsequently recognize the corresponding items (images) in the recognition task. Items that were both correctly responded to during the instruction phase and accurately identified during the recognition task were classified as hits. For each emotional valence, hit rates were calculated as the proportion of correctly recognized instructed items relative to the total number of Recall or Suppression items. False alarm rates were calculated as the proportion of incorrect responses to foils relative to the total number of foils. Sensitivity scores were then computed separately for Recall and Suppression conditions and correlated with BIS values (Figure S2). No associations were observed between BIS and sensitivity scores (all uncorrected one-tailed $p$s > 0.05) indicating that higher BIS values reflected reduced interference from memory representations on the working memory task which is consistent with more effective suppression or less effective recall, rather than differences in recognition sensitivity.

**Figure S2**

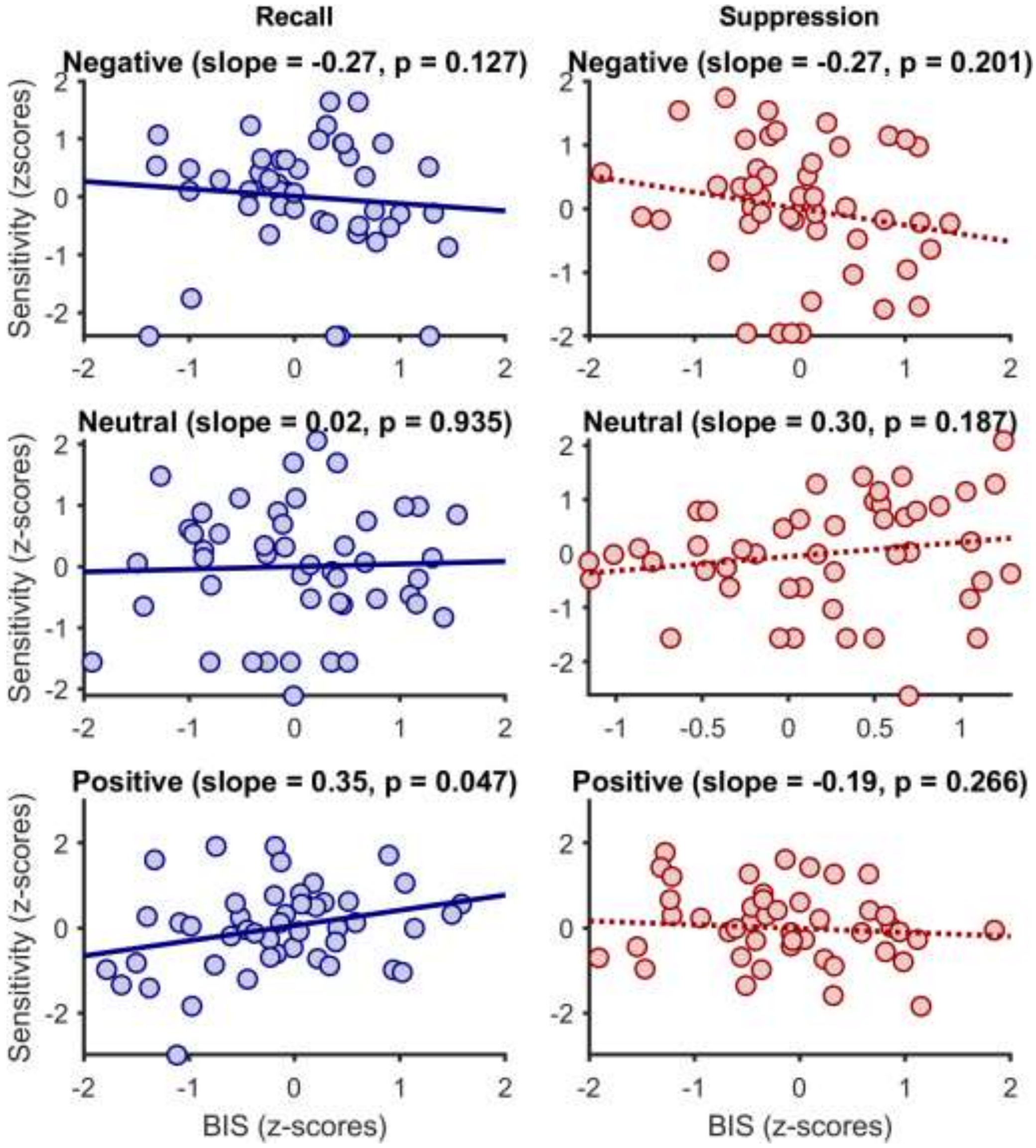

**No association observed between BIS and sensitivity.** The left panels (A) show scatter plots BIS and sensitivity scores (blue markers and solid fitted least square lines) in the Recall condition, while the right panels (B) show BIS and sensitivity scores (red markers and broken fitted least squares lines) in the Suppression condition. Each marker represents one participant. The slope and *p*-values were computed using the MATLAB function – *robustfit*.

**C) Reliability:** To assess the stability of the BIS metric over time and across conditions, we computed split-half correlations between odd and even trials separately for each condition and emotion and applied the Spearman–Brown correction after computing mean across emotions.

All trials were concatenated and subsequently categorized into three emotional conditions separately for Recall and Suppression for each participant. For each emotion, trial-wise accuracy and reaction times were extracted. Trials were then divided into odd and even subsets to form two halves. For each half, accuracy and mean reaction time for correct trials were computed for each emotion and participant. These measures were standardized (z-scored) across participants for each emotion. Split-half reliability was estimated by computing Pearson's correlations between the two halves for each emotion, followed by Spearman–Brown correction (Equation 2).

Spearman-Brown correction = $\frac{2r}{1+r}$ ... (Equation 2)

Mean reliability was then calculated across emotions separately for Recall and Suppression. The resulting mean reliability coefficients were 0.80 for Recall and 0.76 for Suppression (see Figure S3), indicating stability of the metric. The Spearman–Brown reliability coefficient greater than 0.60 is considered acceptable, whereas coefficients between 0.80 and 0.90 are indicative of good reliability (Royal, 2017; Vet et al., 2013).

83

84

85 **Figure S3**

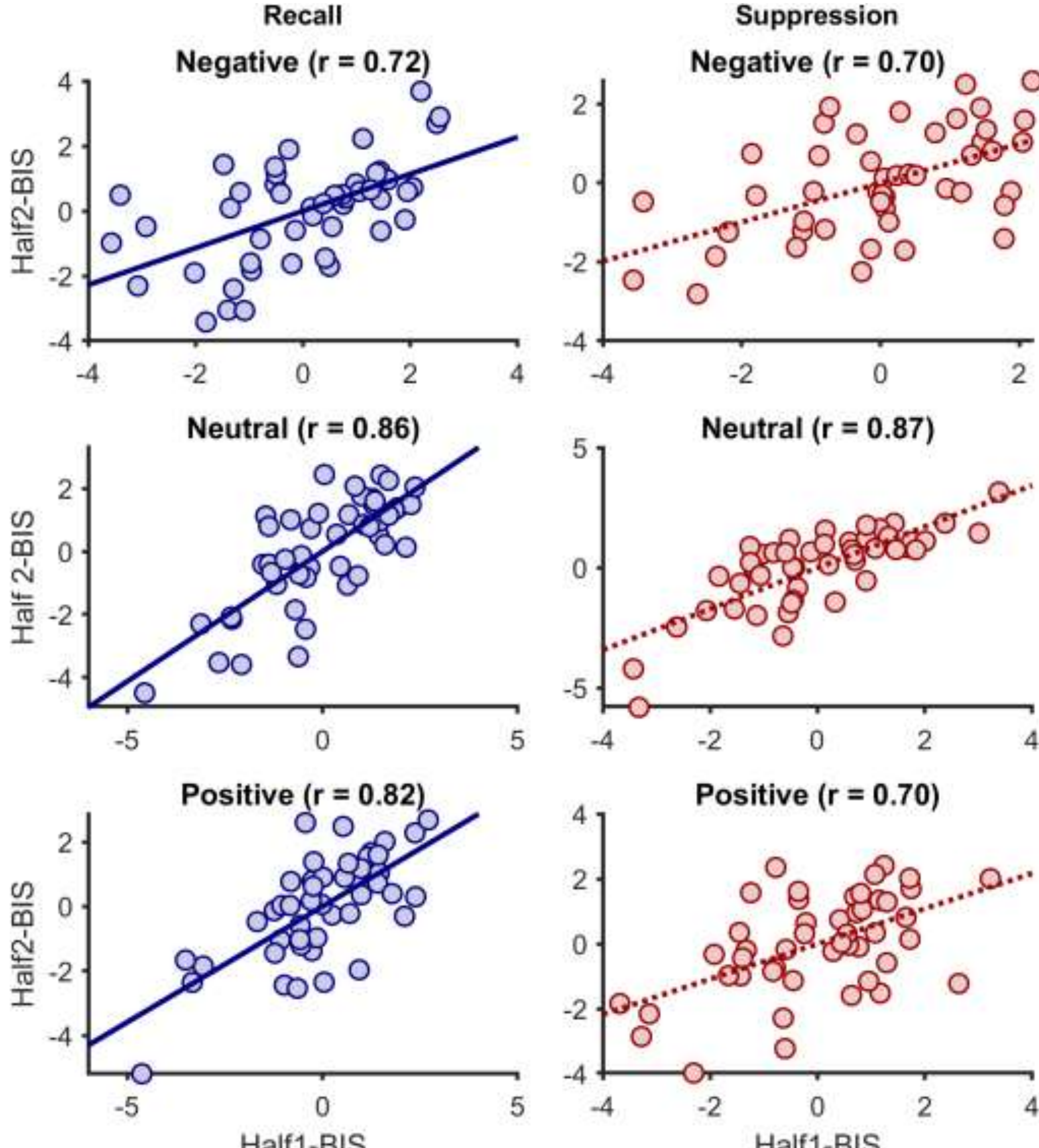

**Reliability of the BIS metric.** The left panels (A) display scatter plots comparing the first half of the BIS metric
with the second half (blue markers with solid least-squares fit lines) for the Recall condition. The right panels (B)
present the same comparison (red markers with dashed least-squares fit lines) for the Suppression condition.
Each marker corresponds to an individual participant. The coefficient of robust regression (r) was computed
using MATLAB function – *robustfit*.

**D) Separate RT and Accuracy analyses -**

To validate the BIS metric, we conducted separate analyses of accuracy and reaction time (RT) for each condition. Accuracy, defined as the proportion of correct responses per condition for each participant, was calculated (Figure S4). A Friedman test revealed no significant effect of condition on accuracy, $\chi^2$ (5) = 6.14, $p$ = 0.29, indicating that accuracy did not differ across the six experimental conditions. Similarly, mean RTs for correct trials were computed for each condition and participant (Figure S5). Again, a Friedman test showed no significant differences across conditions, $\chi^2$ (5) = 8.92, $p$ = 0.11. Together, these findings suggest that dual-task memory-related interference did not manifest in either accuracy or reaction time across conditions.

We also checked the relationship between BIS and accuracy / reaction time by computing Pearson's $r$ and observed a positive trend between BIS and accuracy (Figure S6) and a negative trend between BIS and reaction time (Figure S7). The trend of the relationship suggests that higher accuracy and lower (faster) reaction times is linked with higher BIS or reduced interference, thus better performance in the visual working memory task.

As working memory capacity correlates with BIS, we also examined its relationship with accuracy and reaction time to determine whether these trends were comparable to those observed between BIS and performance measures. Consistent with our expectations, we observed a positive association between working memory capacity and accuracy (Figure S8) and a negative association between working memory capacity and reaction time (Figure S9).

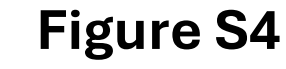

**Figure S4**

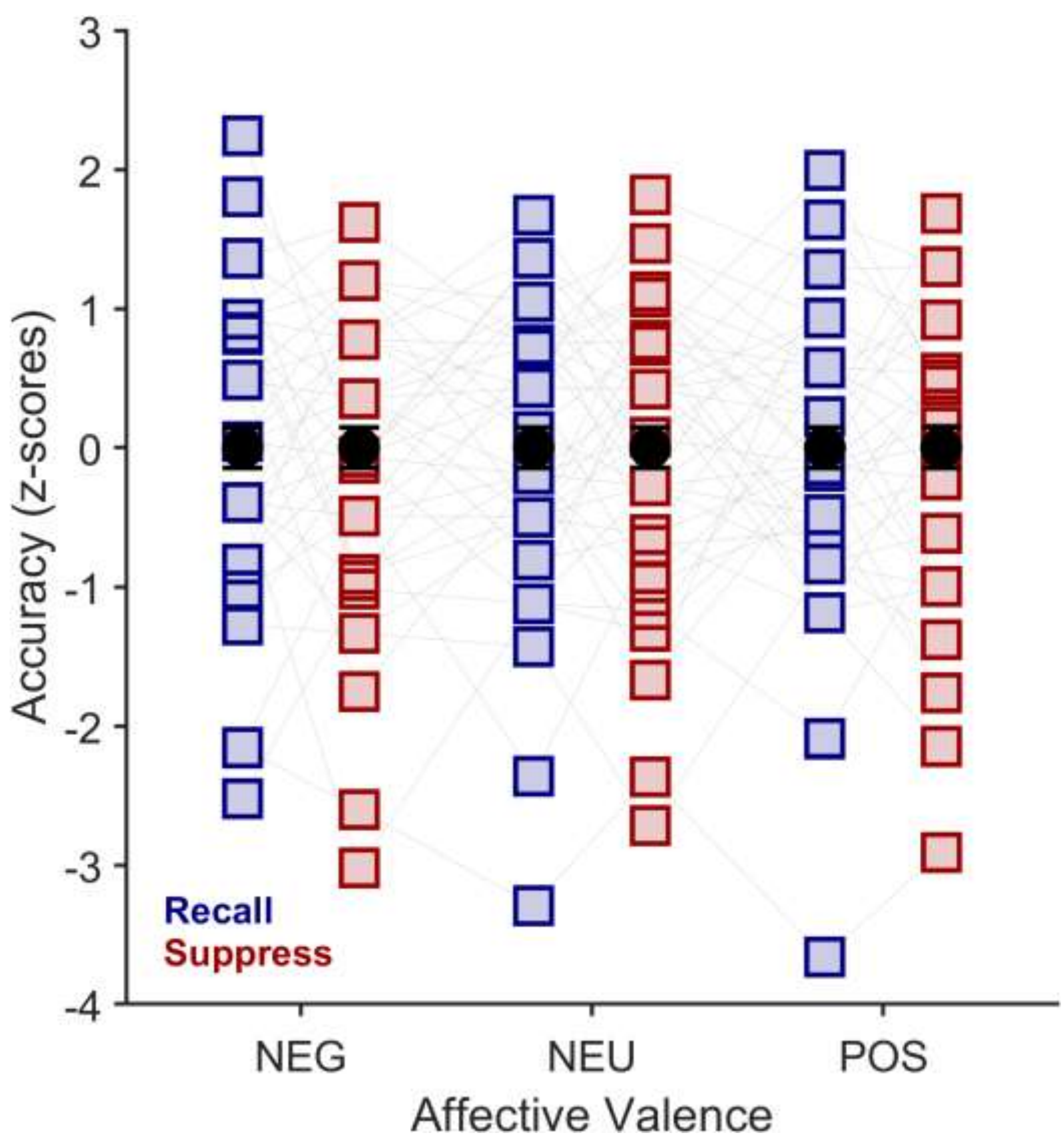

**Effects of task instructions (Recall and Suppress) and/or emotions (negative, neutral, and positive) on accuracy**. (A) Plot showing accuracy of each participant linked with Recall condition and Suppress conditions in blue and red square respectively and joined with grey lines, mean across participants in black circle, and standard error of the mean as error bars in black.

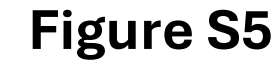
**Figure S5**

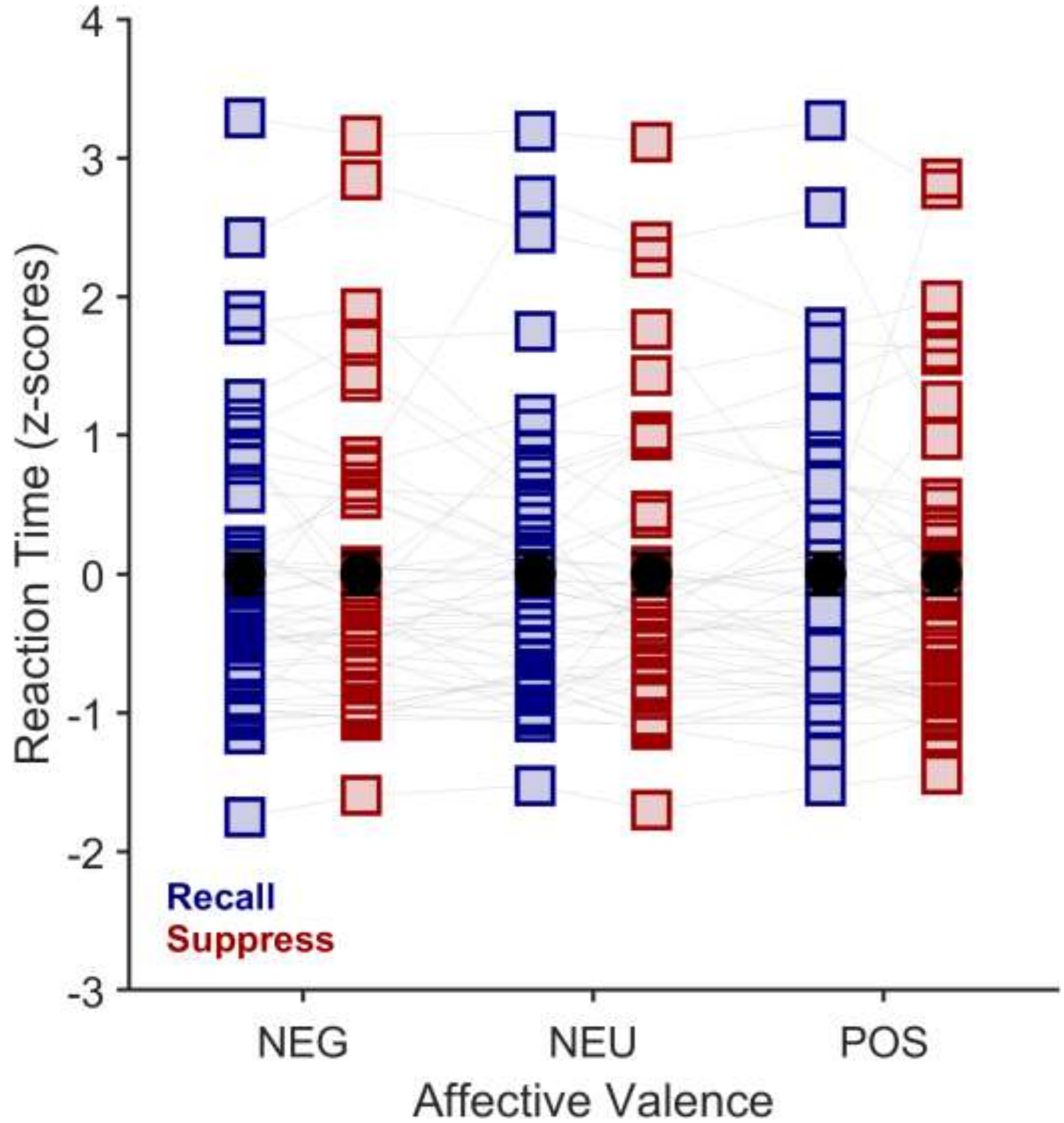

**Effects of task instructions (Recall and Suppress) and/or emotions (negative, neutral, and positive) on reaction time**. (A) Plot showing reaction time of each participant linked with Recall condition and Suppress conditions in blue and red square respectively and joined with grey lines, mean across participants in black circle, and standard error of the mean as error bars in black.

**Figure S6**

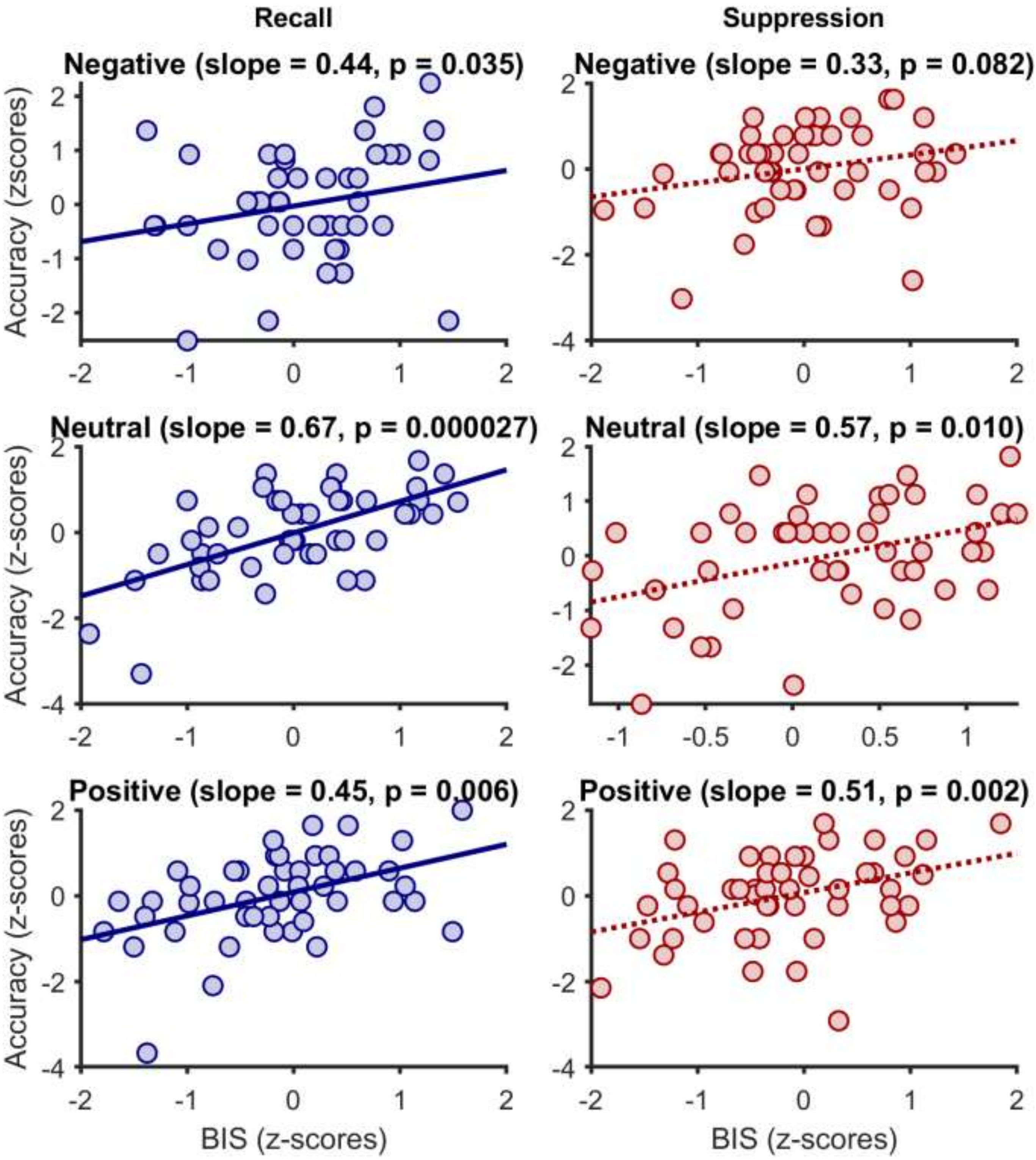

**Trend between BIS and accuracy.** The left panels (A) show scatter plots between BIS and accuracy scores (blue markers and solid fitted least square lines) in the Recall condition, while the right panels (B) show BIS and sensitivity scores (red markers and broken fitted least squares lines) in the Suppression condition. Each marker represents one participant. The slope and *p*-values were computed using the MATLAB function – *robustfit*.

135 **Figure S7**

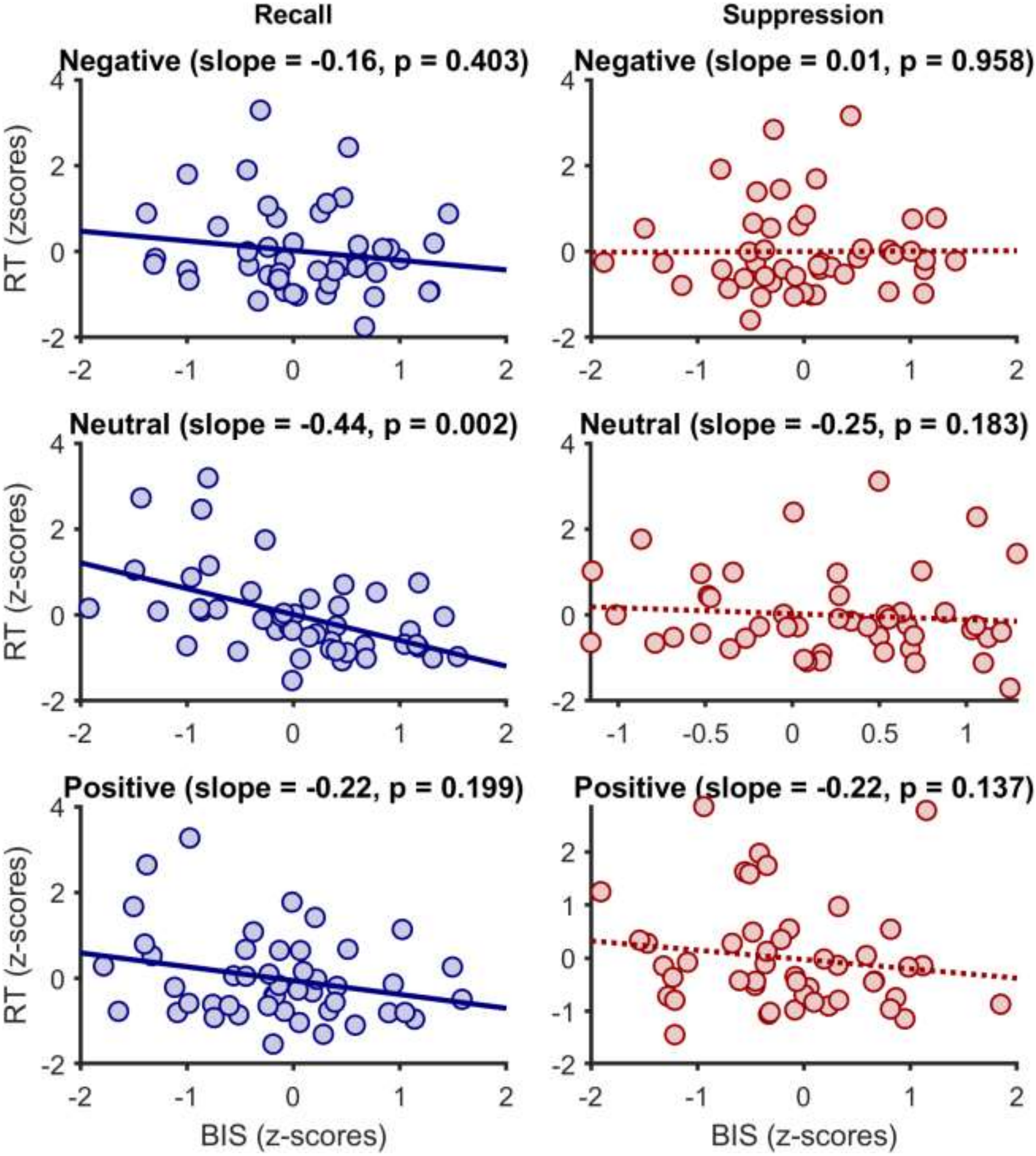

**Trend between BIS and reaction time (RT).** The left panels (A) show scatter plots between BIS and reaction time
scores (blue markers and solid fitted least square lines) in the Recall condition, while the right panels (B) show
BIS and sensitivity scores (red markers and broken fitted least squares lines) in the Suppression condition. Each
marker represents one participant.

**Figure S8**

Recall

Suppression

Negative (slope = 3.41)

Negative (slope = 3.27)

Neutral (slope = 2.40)

Neutral (slope = 2.70)

Positive (slope = 2.74)

Positive (slope = 2.97)

Accuracy (z-scores)

WM capacity (z-scores)

**Trend between working memory (WM) capacity and accuracy.** The left panels (A) show scatter plots between WM capacity and accuracy scores (blue markers and solid fitted least square lines), while the right panels (B) show WM capacity and accuracy scores (red markers and broken fitted least squares lines). Each marker represents one participant. The coefficient of robust regression (r) was computed using MATLAB function – *robustfit*.

**Figure S9**

**Trend between working memory (WM) capacity and reaction time (RT).** The left panels (A) show scatter plots between WM capacity and RT scores (blue markers and solid fitted least square lines), while the right panels (B) show WM capacity and RT scores (red markers and broken fitted least squares lines). Each marker represents one participant. The slope and *p*-values were computed using the MATLAB function – *robustfit*.

### **S2: *Anxiety-Moderated Associations Between Emotional Memory Suppression-linked Cognitive Control and Resting-State Functional Connectivity***

The interaction between anxiety and Suppression-linked cognitive control for negative memories showed a negative association between the left superior frontal gyrus seed and a cluster in the left superior lateral occipital cortex (sLOC l; $T_{(41)} = -5.23$, $p$FWE = 0.026, cluster size $kE = 230$ with center at x = -22, y = -78, z = +50; Figure S3-A). However, this effect did not survive the more stringent threshold of $p$FWE < 0.017. Two separate visualizations revealed a negative correlation in the high (Figure S3-B: $r_{(21)} = -0.60$, $\beta = -0.07$, 95% *CI* [-0.81, -0.25]) and positive correlation in the low (Figure S3-B: $r_{(22)} = 0.46$, $\beta = 0.11$, 95% *CI* [0.06, 0.73]) subset between the Suppression-linked cognitive control for negative memories or BIS and seed-to-voxel rsFC. Closer inspection of the data confirmed the difference between the two slopes, which suggests that the rate of change in rsFC with unit increment of BIS was higher in the low anxiety subset as compared to the high anxiety subset.

**Figure S10**

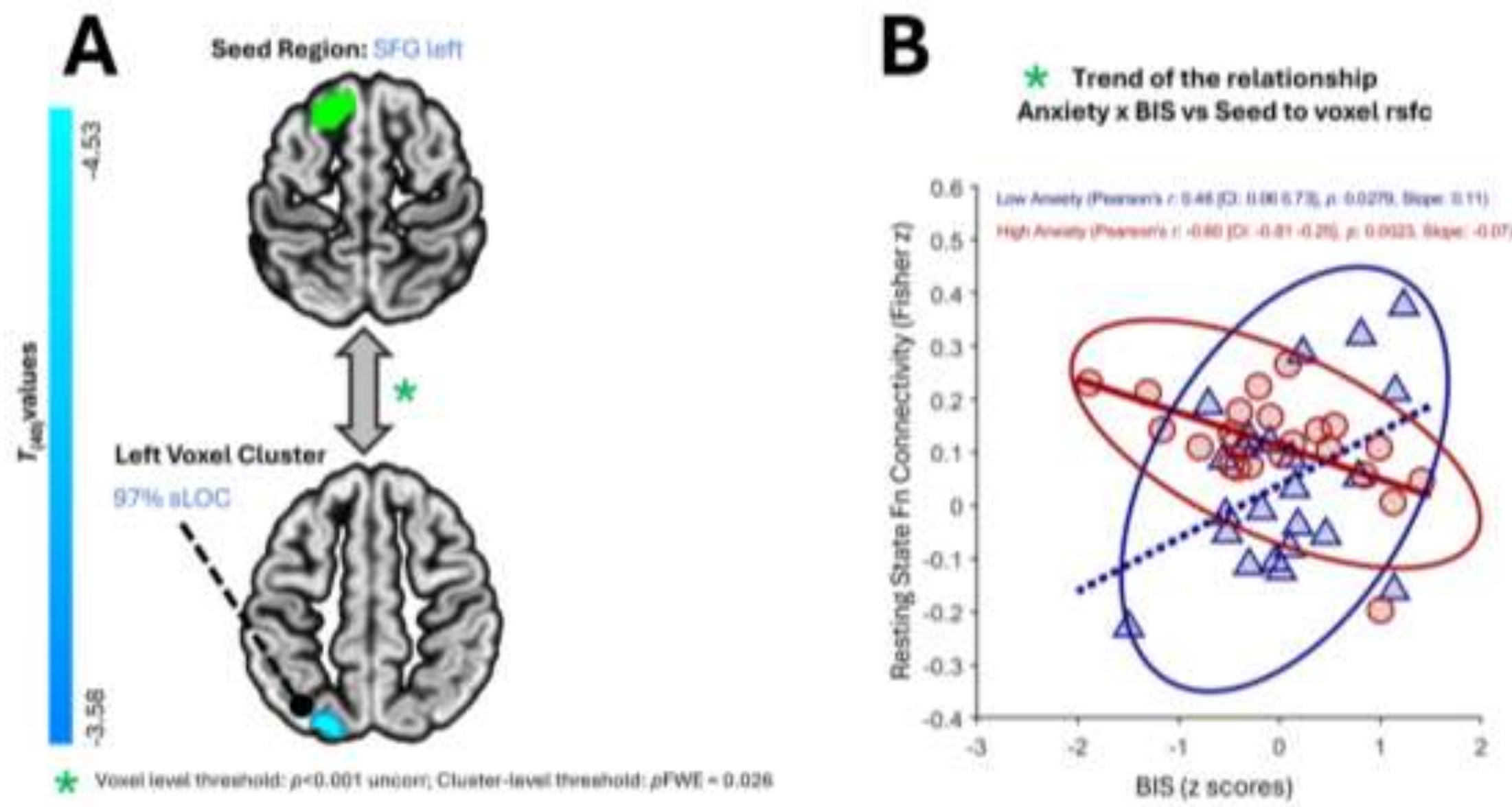

**Moderating effect of anxiety on the association between Suppression-linked Cognitive Control or BIS and rsFC.** (A, C) rsFC between seed region and the cluster. The colourbar depicts the direction and strength of connectivity. The cooler colourbars (A, C) depict negative correlation. (B, D) Scatter plots of the associations between BIS and rsFC shown in A and C respectively. Each marker depicts one participant, and least square lines show the trend in the associations in high (red circle and solid line) and low (blue triangle and dotted line) anxiety subset. The confidence ellipses represent 95% confidence interval. SFG: Superior Frontal Gyrus, sLOC: superior division, Lateral Occipital Cortex ***pFWE < 0.05**

178 **Supplementary Table 1 attached below**

179

| **Analysis** | **Seed** | **Clusters** | **MNI-coordinates**<br><br>**(X Y Z)** | **Cluster size** | **Cluster level *p*-FWE** | **Effect size (T-value)** |
|---|---|---|---|---|---|---|
| Suppression of Negative memories X FA vs rsFC regression | SFG l<br>Superior Frontal Gyrus<br>left | -22 -78 +50<br><br>sLOC l Lateral Occipital Cortex superior division left | <br><br>-22 -78 +50 | 230<br><br>225 | 0.026 | $T_{(40)} = -5.23$ |

180

181

182 **MRI results:** Apriori regions that survived multiple correction (*p*FWE ≤ 0.015)

183